\documentclass[a4paper,11pt]{article}
\pdfoutput=1 

\usepackage{jheppub} 
\makeatletter
\DeclareRobustCommand*{\bfseries}{%
  \not@math@alphabet\bfseries\mathbf
  \fontseries\bfdefault\selectfont
  \boldmath
}
\makeatother

\usepackage{calc}
\usepackage{rotating}
\usepackage[english]{babel}
\usepackage[utf8]{inputenc}
\usepackage{graphicx}
\usepackage{subfig}
\usepackage{float}
\usepackage{amsmath}
\usepackage{amssymb}
\usepackage{amsthm}
\usepackage{slashed}
\usepackage{latexsym}
\usepackage{dcolumn}
\usepackage{hyperref}
\usepackage{color}
\usepackage{ulem}
\usepackage[table,xcdraw,dvipsnames]{xcolor}
\usepackage{placeins}

\def\mv{m_V}

\def\mmu{M_\mu}

\def\mBs{M_{B_s}}

\def\dgmtwo{\delta(g-2)_{\mu}}

\def\O{\mathcal{Q}}
\def\Ot{\widetilde{\mathcal{Q}}}

\def\deft{\textrm{DEFT}}

\def\C{\mathcal{C}}
\def\Ct{\widetilde{\mathcal{C}}}

\newcommand{\amc}{{\sc MadGraph5}\_a{\sc MC@NLO}}
\newcommand{\fr}{{\sc Feyn\-Rules}}

\restylefloat{figure}

\preprint{TTP21-020, P3H-21-047}
\title{Flavour anomalies and the muon \texorpdfstring{$\boldsymbol{g-2}$}{g-2} from feebly interacting particles}

\author[a]{Luc Darm\'e,}
\author[b]{Marco Fedele,}
\author[c]{Kamila Kowalska}
\author[c]{and Enrico Maria Sessolo}

\affiliation{$^a$ INFN, Laboratori Nazionali di Frascati,\\ C.P. 13, 100044 Frascati, Italy\\
$^b$ Institut f\"ur Theoretische Teilchenphysik, Karlsruhe Institute of Technology,\\ D-76131 Karlsruhe, Germany\\
$^c$ National Centre for Nuclear Research,\\ ul.~Pasteura~7, 02-093 Warsaw, Poland}

\emailAdd{luc.darme@lnf.infn.it}
\emailAdd{marco.fedele@kit.edu}
\emailAdd{kamila.kowalska@ncbj.gov.pl}
\emailAdd{enrico.sessolo@ncbj.gov.pl}

\abstract{We perform a phenomenological 
analysis of simplified models of light, feebly interacting particles~(FIPs)
that can provide a 
combined explanation of the anomalies in $b\to s l^+ l ^-$ transitions at LHCb and
the anomalous magnetic moment of the muon. 
Different scenarios are categorised according to the explicit momentum dependence of the FIP coupling to the $b-s$ and $\mu-\mu$ vector currents and they are 
subject to several constraints from flavour and precision physics. We show that viable combined solutions to the muon $g-2$ and flavour anomalies exist  with the exchange of a vector FIP with mass larger than $4 \,\textrm{GeV}$.
Interestingly, the LHC has the potential to probe this region of the parameter space by increasing the precision of the $Z\to 4\mu$ cross-section measurement. Conversely, we find that solutions based on the exchange of a lighter vector, in the $m_V < 1\,\textrm{GeV}$ range, are essentially excluded by a combination of $B\to K +\textrm{invisible}$ and $W$-decay precision bounds. }

\begin{document}
\maketitle
\section{Introduction\label{sec:intro}}

Feebly interacting particles~(FIPs) represent a very large and well-motivated category of new physics~(NP) scenarios. Loosely defined as new light particles with mass below the electroweak symmetry-breaking (EWSB) scale and with a feeble interaction with Standard Model~(SM) fields, FIPs encompass NP particles as diverse as the dark photon and axion-like particles.
A particularly exciting possibility~\cite{Datta:2017pfz,Sala:2017ihs,Datta:2017ezo,Alok:2017sui,Altmannshofer:2017bsz,Datta:2018xty,Datta:2019bzu,Darme:2020hpo} is that one or more FIPs may be responsible for the anomalies in lepton-flavour universality violating~(LFUV) observables recently confirmed 
in new data from LHCb
and for the discrepancy between the measured value of the anomalous magnetic moment of the muon and 
its SM expectation. 

The LHCb Collaboration recently released an updated measurement of the ratio $R_K=\textrm{BR}(B\to K \mu^+ \mu^-)/\textrm{BR}(B\to K e^+ e^-)$, which included the full Run~I + Run~II data sets. The current value deviates from the SM prediction by more than $3\,\sigma$~\cite{Aaij:2021vac}. Once the LFUV ratio $R_{K^{\ast}}=\textrm{BR}(B\to K^{\ast} \mu^+ \mu^-)/\textrm{BR}(B\to K^{\ast} e^+ e^-)$~\cite{Aaij:2017vbb,Belle:2019oag} and the branching ratios and angular observables of other decays mediated by $b\to s\, l^+ l^-$ transitions~\cite{LHCB:2014cxe,Aaij:2015dea,Aaij:2015esa,Khachatryan:2015isa,Aaij:2015oid,Aaij:2016flj,Wehle:2016yoi,Sirunyan:2017dhj,Aaboud:2018krd,Aaij:2020nrf,Aaij:2020ruw} are considered as well, a global picture emerges, pointing to the potential 
presence of NP interacting with the muons. These contributions are statistically favoured compared to the SM prediction alone, 
at the level of more than $5\,\sigma$, see Refs.~\cite{DAmico:2017mtc,Ciuchini:2019usw,Alguero:2019ptt,Alok:2019ufo,Datta:2019zca,Aebischer:2019mlg,Kowalska:2019ley,Ciuchini:2020gvn,Hurth:2020ehu,Alda:2020okk,Altmannshofer:2021qrr,Geng:2021nhg,Alguero:2021anc,Ciuchini:2021smi} for recent analyses. 

Equally intriguing is the recent measurement of the anomalous magnetic moment of the muon by the 
E989 experiment at Fermilab~\cite{PhysRevLett.126.141801}. The experimental collaboration reports a $3.3\,\sigma$ deviation from the value expected in the 
SM~\cite{Davier:2017zfy,Keshavarzi:2018mgv,Colangelo:2018mtw,Hoferichter:2019mqg,Davier:2019can,Keshavarzi:2019abf,Kurz:2014wya,Melnikov:2003xd,Masjuan:2017tvw,Colangelo:2017fiz,Hoferichter:2018kwz,Gerardin:2019vio,Bijnens:2019ghy,Colangelo:2019uex,Colangelo:2014qya,Blum:2019ugy,Aoyama:2012wk,atoms7010028,Czarnecki:2002nt,Gnendiger:2013pva}. When the new measurement is statistically combined with the previous experimental determination, obtained a couple of decades ago at Brookhaven~\cite{Bennett:2006fi}, one gets~\cite{PhysRevLett.126.141801}
\begin{equation}\label{eq:meas}
\delta(g-2)_{\mu}=\left(2.51\pm 0.59 \right)\times 10^{-9}\,,
\end{equation}
which yields a deviation from the SM data-driven prediction at the $4.2\,\sigma$ level. 

If the LFUV anomalies and $\delta(g-2)_{\mu}$ 
are due to the interactions of one or more FIPs in the MeV to GeV range, one expects the 
accompanying presence of new flavoured physics around the TeV scale (see, e.g., Ref.~\cite{Darme:2020hpo}). On the one hand, higher-dimensional FIP interactions, such as the ones characteristic of axion-like particle models, are simply effective field theories (EFTs) requiring an ultraviolet (UV) completion. On the other hand, even for renormalisable FIP interactions the presence of NP above the EWSB scale is often required, for instance to evade the strong bounds from neutrino trident production, which otherwise drastically constrains a solution to the $(g-2)_{\mu}$ anomaly based on light states~\cite{Altmannshofer:2014pba}.

While a model-independent description of heavy physics in flavour-violating meson decay via the Weak Effective Theory (WET) has proved invaluable in identifying the specific operators associated with the anomalies emerging in a large set of observables, the WET fails to account properly for the possibility of one or more \textit{extra} degrees of freedom that are light but hidden, as it is not by construction equipped to take into account the effects of momentum-dependent couplings, or the presence of possible resonances in the experimental energy bins. We thus consider in this work EFT operators more suited to the study of $b \to s $ and $\mu$-related physics under the assumption that the unspecified heavy NP is accompanied by a light FIP~\cite{Altmannshofer:2017yso,Arina:2021nqi} (see also dark-matter motivated constructions, for instance Refs.~\cite{Fox:2011fx,Duch:2014xda,Bishara:2016hek,DeSimone:2016fbz,Darme:2020ral}). 

To this end, we construct a set of operators that parametrise the interactions of the FIP with Lorentz-invariant vector bilinears of the $b-s$ and $\mu-\mu$ current, in agreement with the measurement of the LFUV and $(g-2)_{\mu}$ anomalies. They are characterised by increasing powers of the FIP momentum transferred in the low-energy process. We then perform a comprehensive analysis of the constraints that can be applied on the Wilson coefficients of these new operators. Constraints arise from several sources: direct measurements of the branching ratios for $B\to K +\textrm{invisible}$ and $B\to K^{\ast}\mu^+\mu^-$ resonant decays, the measurements of 
$\textrm{BR}(B_s\to \mu^+\mu^-)$ and $B_s$-mixing, and several precision measurements associated
with the $W$- and $Z$-boson decay widths. 

The purpose of our analysis is twofold. Firstly, we expand significantly on our previous study of a light ``dark'' U(1)$_D$ gauge boson solution to the $b-s$ anomalies with a $q^2$-dependent coupling~\cite{Darme:2020hpo}. On the one hand, we add strong new constraints that were not considered previously in the literature (we compute numerically the best currently available bounds on GeV-scale muonphilic particles from the recent ATLAS analysis of $Z\to 4\mu$~\cite{ATLAS:2021kog} and the limits based on the violation of flavour universality in $W$-boson decay~\cite{Zyla:2020zbs}). On the other, we perform a global scan of two new simplified models with different momentum dependence of the couplings with respect to Ref.~\cite{Darme:2020hpo}. One of these models, incidentally, overlaps with the case considered in Ref.~\cite{Sala:2017ihs}, for which we exclude the region of parameter space highlighted there and point to a slightly heavier preferred solution for the flavour and $(g-2)_{\mu}$ anomalies, with no fine tuning of the vector and axial-vector $Z'$ couplings to the muon. The second and main goal of this work is to provide a concise and broad model-independent picture of the current phenomenological status of a light-FIP solution to the muon anomalies. In the process we will establish a ``dictionary'' allowing the reader to map easily the Wilson coefficients of a FIP-friendly EFT to the WET Wilson coefficients used in the publicly available numerical packages. 

The paper is organised as follows. In Sec.~\ref{sec:operators} we introduce the operators better suited to describing FIP interactions in $b\to s$ and LFUV processes. In Sec.~\ref{sec:match} we provide the correspondence with the WET basis usually employed in analyses of rare meson decays. The list of applied constraints is introduced in Sec.~\ref{sec:flav-phys}. The result of a global scan in the new operator basis are presented and discussed in Sec.~\ref{sec:numerics}, and we conclude in Sec.~\ref{sec:summary}. In Appendix~\ref{app:a} we show the details required to perform a full calculation of processes involving the $B$ meson in the basis introduced in Sec.~\ref{sec:operators}.

\section{EFT plus light degrees of freedom\label{sec:operators}}

We introduce in this section Dark EFT (DEFT) operators of the Lagrangian connecting a generic FIP to the (axial-)vector SM currents relevant for the LFUV and $(g-2)_{\mu}$ anomalies. 

\paragraph{Spin-1 FIP} We parametrise the exchange of a light vector $V$ out of the left-handed $b-s$ current,  in agreement with the results of the global fits, in terms of the following operators~\cite{Altmannshofer:2017bsz}:
\begin{align}
&\O^{bsV}_4=\left(\bar{s}\gamma_{\rho}P_L b\right) V^{\rho}\,,\label{eq:bs_oper_4} \\
&\O^{bsV}_6=\left(\bar{s}\gamma_{\rho}P_L b\right)\partial_{\sigma}V^{\rho\sigma}\,,\label{eq:bs_oper_6} 
\end{align}
where we have used the field strength tensor $V^{\rho\sigma}=\partial^{\rho} V^{\sigma}-\partial^{\sigma} V^{\rho}$.\footnote{We leave for future work the dimension~$5$ case, e.g., $\O^{bsV}_5=\left(\bar{s}\sigma_{\rho \sigma}P_R b\right)V^{\rho\sigma}$.
Since it couples the FIP to the tensor SM current it leads to a starkly different phenomenology.} 
We complete Eqs.~(\ref{eq:bs_oper_4}), (\ref{eq:bs_oper_6}) with the corresponding interactions with the muon current, 
\begin{align}
&\O^{\mu\mu V}_4=\left(\bar{\mu}\gamma_{\rho} \mu \right) V^{\rho}\,,&& \Ot^{\mu\mu V}_4=\left(\bar{\mu}\gamma_{\rho} \gamma^5 \mu \right) V^{\rho}\,, \label{eq:mu_oper_4} \\
&\O^{\mu\mu V}_6=\left(\bar{\mu }\gamma_{\rho}  \mu \right)\partial_{\sigma}V^{\rho\sigma}\,, && \Ot^{\mu\mu V}_6=\left(\bar{\mu}\gamma_{\rho} \gamma^5 \mu \right) \partial_{\sigma}V^{\rho\sigma}\,. \label{eq:mu_oper_6}
\end{align}
Since we do not specify the origin of these interactions, Eqs.~(\ref{eq:bs_oper_4})-(\ref{eq:mu_oper_6}) will serve in describing the $B$-physics for all light vector states. These include, in particular, the GeV-scale top-philic particle in Ref.~\cite{Fox:2018ldq} and the renormalisable model introduced in Ref.~\cite{Darme:2020hpo}.

\paragraph{Spin-0 FIP: the pNGB}

We consider next the case of a (pseudo-)scalar FIP, $a$, possibly a pseudo Nambu-Goldstone boson (pNGB). The standard derivative interaction term with the quark current reads 
\begin{align}
 &\O^{b s a}_5=   \frac{1}{2}\, \partial_{\rho}a\,\left(\bar{s}\gamma^{\rho}  P_L b \right)\,,\label{eq:pNGB_bs}
\end{align}
and 
\begin{align}
 &\O^{\mu\mu a}_5=   \frac{1}{2}\,\partial_{\rho}a\,\left(\bar{\mu}\gamma^{\rho}  \mu\right),  & \quad & \Ot^{\mu\mu a}_5=   \frac{1}{2}\, \partial_{\rho}a\, \left(\bar{\mu}\gamma^{\rho} \gamma^5 \mu\right), \label{eq:pNGB_mu}
\end{align}
are the corresponding couplings to the muon current. These interactions are strongly constrained by flavour physics and, more importantly for our purposes, do not lead to the vector four-fermion operators $\mathcal{O}_9^{\mu(\prime)}$ and $\mathcal{O}_{10}^{\mu(\prime)}$ of the WET, which induce the solution to the LFUV anomalies preferred in the global scans. 
We therefore leave the study of these operators for future work.

\paragraph{FIPs as dark matter: spin 0, 1/2}

When the light state is a stable new particle, which can play the role of dark matter, one needs to include operators with two FIP insertions at dimension 6. Using the dark current,
\begin{align}
    \label{eq:darkcurrent}
    \mathcal{J}_{D}^{\rho} = \begin{cases}
    i \left( S^* \partial^\rho S - S \partial^\rho S^* \right)\,\quad (\textrm{scalar, }D=S) \\
 \bar \chi \gamma^\rho \chi\,, \quad \quad \quad  \quad \quad \quad  (\textrm{fermion, }D=\chi)
    \end{cases} 
\end{align}
we can then define
\begin{align}
 &\O^{ \mu\mu \chi \chi }_6=  \mathcal{J}_{\chi}^{\rho} \left(\bar{\mu}\gamma_{\rho}  \mu \right)\,,  & & \Ot^{\mu\mu \chi \chi}_6=    \mathcal{J}_{\chi}^{\rho}  \left(\bar{\mu}\gamma_{\rho} \gamma^5 \mu \right)\,, \label{eq:dm_oper_1}
\end{align}
and 
\begin{align}
 &\O^{b s \chi \chi}_6 =  \mathcal{J}_{\chi}^{\rho} \left( \bar{s}\, \gamma_{\rho}  P_L b\right)  \,, \label{eq:dm_oper_2}
\end{align}
and equivalent operators can be defined in the scalar case, with $\chi \to S$.

These effective interactions can be used to parametrise, for example, UV completions to the WET operators $\mathcal{O}_9^{\mu(\prime)}$, $\mathcal{O}_{10}^{\mu(\prime)}$ 
based on dark-matter induced box loops, in the case where the dark matter is relatively light. Note, however, that in such a minimal scenario the dark matter mass is expected to be large enough to evade strong limits from the direct decay $B\to K +\textrm{invisible}$~\cite{Darme:2020hpo}.

\section{Relation with the WET\label{sec:match}}

In order to quantify the effects of DEFT operators on the flavour observables, 
we first create a dictionary between the DEFT and the WET.
The procedure is equivalent to considering an explicit momentum dependence of the WET Wilson coefficients, of the type explored, e.g., in Refs.~\cite{Datta:2017ezo,Datta:2017pfz,Datta:2018xty,Darme:2020hpo}. 
The specific Lorentz structure of the DEFT operators considered in this work allows us 
to draw the correspondence with the WET straightforwardly at the level of the spinor fields. We show in Appendix~\ref{app:a} 
that the results of this section agree with a full calculation of $B$-meson decays.

\subsection{Tree level\label{sec:tree-mat}}

\paragraph{Spin-0 FIP} The critical ingredient in linking both theories is that we are interested in physical processes in which the muons are on-shell. Given two on-shell external muons of four-momenta $k_1, k_2$, emerging from the exchange of a FIP of momentum $q = k_1 + k_2$, 
the equations of motions yield
\begin{eqnarray}
     q^\rho \left(\bar \mu \gamma_\rho \mu \right) & =  & 0 \\
     q^\rho \left(\bar \mu \gamma_\rho \gamma^5 \mu \right) & = & 2 M_\mu \left(\bar \mu \gamma^5 \mu \right),
\end{eqnarray}
where $M_{\mu}$ is the muon mass. A direct consequence of the above equations is that the derivative scalar current introduced in Eqs.~(\ref{eq:pNGB_bs}) and (\ref{eq:pNGB_mu}) will give rise to the standard pseudoscalar interactions $\mathcal{O}_P^{\mu(\prime)}$, and thus cannot generate the WET vector operators $\mathcal{O}_9^{\mu(\prime)}$ and $\mathcal{O}_{10}^{\mu(\prime)}$. 

By repeating the argument for ``on-shell'' initial quarks with $q=p_b-p_s$, one obtains
\begin{align}\label{eq:smom}
    & \left(\bar s \gamma_\rho  P_L b \right) q^\rho  =   M_s \left(\bar{s}P_L b \right)-M_b \left(\bar{s}P_R b \right),
\end{align}
where $M_{s(b)}$ is the strange (bottom) quark mass. 
The above equality is useful for deriving the correspondence between DEFT and WET in the remainder of this section.

\paragraph{Spin-1 FIP} We write the vector FIP propagator in the form
\begin{equation}\label{eq:pi}
    \Pi^{\rho \sigma} = \frac{-i\left(g^{\rho \sigma} - q^\rho q^\sigma / m_V^2\right)}{\Pi}\,,\quad
    \textrm{with}\quad \Pi = q^2 - m_V^2 + i \Gamma_V \sqrt{q^2}\,,
\end{equation}
where $m_V$ and $\Gamma_V$ indicate, respectively, the mass and total decay width of the light NP 
vector $V^{\rho}$. By considering the $ b \to s \mu \mu$ process via a vector FIP exchange, we can 
derive the relation between the two effective theories. In the case, for example, of an 
interaction involving operators $\O_6^{bsV}$ and $\Ot_4^{\mu \mu V}$ in Eqs.~(\ref{eq:bs_oper_6}) and (\ref{eq:mu_oper_4}) one finds the amplitude 
\begin{align}\label{eq:64composition}
\mathcal{M}&=-\, \Ct_4^{\mu\mu V} \C_6^{bsV}/\Lambda^2 \left(\bar{s}\gamma_{\rho}P_L b \right)
\left(q^2 g^{\sigma\rho}-q^{\sigma}q^{\rho} \right)  \Pi_{\sigma \nu}\left(\bar{\mu}\gamma^{\nu}\gamma^5 \mu \right) \nonumber\\
&= i\frac{ \Ct_4^{\mu\mu V} \C_6^{bsV}/\Lambda^2}{\Pi}
\left(\bar{s}\gamma_{\rho}P_L b \right)\left(q^2\delta^{\rho}_{\nu}-q^{\rho}q_{\nu} \right)\left(\bar{\mu}\gamma^{\nu}\gamma^5 \mu \right).
\end{align}
The above amplitude is 
expressed in terms of the UV cutoff, $\Lambda$, and the DEFT 
dimensionless coefficients $\C_6^{bsV}$, $\Ct_4^{\mu\mu V}$. 
This leads in turn to replacing the weak effective Hamiltonian for $b\to s\mu\mu$ with
\begin{align}
\mathcal{H}&\to -\frac{ \Ct_4^{\mu\mu V} \C_6^{bsV}}{\Lambda^2\,\Pi}\frac{4\pi}{\alpha_{\textrm{em}}}\left[q^2 \mathcal{O}_{10}^{\mu}+2M_{\mu}\left(M_s\mathcal{O}_P^{\mu\prime}-M_b\mathcal{O}_P^{\mu} \right)  \right]\,,
\end{align}
where we have used the relation~(\ref{eq:smom}).

Following a similar procedure, we can connect to the WET all the operators introduced in Sec.~\ref{sec:operators}$a$. 
The WET coefficients $C_9^{\mu}$, $C_{10}^{\mu}$, $C_P^{\mu}$, $C_P^{\mu\prime}$, 
can then be expressed in terms of DEFT coefficients. One gets,
\begin{align}
  C_9^{\mu}  &\to \frac{4\pi\mathcal{N}}{\alpha_{\textrm{em}}} \frac{(\C_4^{bsV} +\frac{q^2}{\Lambda^2} \C_6^{bsV}  )( \C_4^{\mu\mu V} + \frac{q^2}{\Lambda^2} \C_6^{\mu\mu V} )}{q^2 - m_V^2 + i \Gamma_V \sqrt{q^2}}\,, \label{eq:match_c9} \\[0.8em]
   C_{10}^{\mu}  &\to \frac{4\pi\mathcal{N}}{\alpha_{\textrm{em}}}  \frac{(\C_4^{bsV} +\frac{q^2}{\Lambda^2} \C_6^{bsV}  )( \Ct_4^{\mu\mu V} +\frac{q^2}{\Lambda^2} \Ct_6^{\mu\mu V} )}{q^2 - m_V^2 + i \Gamma_V \sqrt{q^2}}\,, \label{eq:match_c10} \\[0.8em]
   C_{P}^{\mu}  &\to -\frac{4\pi\mathcal{N}}{\alpha_{\textrm{em}}}  \frac{\frac{2M_\mu M_b}{m_V^2}\C_4^{bsV}\left(\Ct_4^{\mu\mu V}+\frac{m_V^2}{\Lambda^2}\Ct_6^{\mu\mu V}\right)+\frac{2M_\mu M_b}{\Lambda^2}\C_6^{bsV}\left(\Ct_4^{\mu\mu V}+\frac{q^2}{\Lambda^2}\Ct_6^{\mu\mu V}\right)}{q^2 - m_V^2 + i \Gamma_V \sqrt{q^2}}\,, \label{eq:match_cp}
   \\[0.8em]
   C_{P}^{\mu\prime}  &\to \frac{4\pi\mathcal{N}}{\alpha_{\textrm{em}}}  \frac{\frac{2M_\mu M_s}{m_V^2}\C_4^{bsV}\left(\Ct_4^{\mu\mu V}+\frac{m_V^2}{\Lambda^2}\Ct_6^{\mu\mu V}\right)+\frac{2M_\mu M_s}{\Lambda^2}\C_6^{bsV}\left(\Ct_4^{\mu\mu V}+\frac{q^2}{\Lambda^2}\Ct_6^{\mu\mu V}\right)}{q^2 - m_V^2 + i \Gamma_V \sqrt{q^2}}\,, \label{eq:match_cpp}
\end{align}
where $\mathcal{N}^{-1}=(4 G_F/\sqrt{2})\,V_{tb}V^{\ast}_{ts}$\,.

\paragraph{An explicit UV example} We conclude this subsection by matching the DEFT operators to an example of renormalisable UV-complete model in the case of a vector FIP. We focus on the flavour-violating coupling to $b$ and $s$.

Let us consider the SM extended by a ``dark'' U(1)$_D$ gauge group, 
whose gauge boson is a light $V$. Let us then introduce a scalar field $\phi$ and a multiplet $\chi$ of vector-like fermions of mass $m_{\chi}$ whose charges with respect to SU(3)$_c\times$SU(2)$_L\times$U(1)$_Y\times$U(1)$_D$ are given by
\begin{equation}\label{eq:np_fields}
    \phi:(\mathbf{1},\mathbf{1},0,Q_{\phi})\, \quad \quad  \chi:(\mathbf{3},\mathbf{2},1/6,Q_{\chi})\,.
\end{equation}
When $Q_{\phi}=Q_{\chi}$ the fields in Eq.~(\ref{eq:np_fields}) admit Yukawa interactions with the SM quark doublets $Q_i=(P_L u_i,P_L d_i)^T$ (of generation $i=2,3$) of the type
\begin{equation}
\mathcal{L}\supset Y^i_Q\, \phi\bar{\chi}\, Q_i+\textrm{H.c.}\label{eq:lagr}
\end{equation}

Depending on whether or not the field $\phi$ acquires a vacuum expectation value~(vev) different DEFT operators of the $b-s$ current are generated at the leading order from the coupling in Eq.~(\ref{eq:lagr}). If $\phi$ develops a vev, $v_{\phi}$, the effective coupling of $V$ to $b-s$, which is generated via the mixing of the SM and NP quarks, does not carry $q^2$ dependence and thus gives rise to the operator $\O^{bsV}_4$. One gets 
\begin{equation}\label{eq:UV_c4}
    \mathcal{C}^{bsV}_4=g_D Q_{\phi}\frac{Y^{s\ast}_Q\, Y^b_Q\, v_{\phi}^2}{2 m_{\chi}^2+ \left( \left|Y^s_Q\right|^2+\left| Y^b_Q\right|^2 \right)v_{\phi}^2}\,,
\end{equation}
where $g_D$ is the dark gauge coupling. The phenomenology of the light gauge boson $V$
in relation to $R_{K^{(\ast)}}$ was analysed, e.g., in Ref.~\cite{Sala:2017ihs}.  

If, on the other hand, the scalar $\phi$ does not develop a vev and U(1)$_D$ is broken by other means unspecified here, the coupling of $V$ to the $b-s$ current is generated
via a penguin diagram constructed out of a $\phi \chi$ loop. Terms presenting an explicit $q^2$ dependence give rise to the operator $\O^{bsV}_6$. Its Wilson coefficient reads 
\begin{equation}\label{eq:UV_c6}
    \frac{\mathcal{C}_6^{bsV}}{\Lambda^2}=\frac{g_D\, Q_{\phi} Y^{s\ast}_Q\, Y^b_Q }{16 \pi^2 m_{\chi}^2} \mathcal{F}\left( \frac{m_{\phi}^2}{m_{\chi}^2}\right)\,,
\end{equation}
where $m_{\phi}$ is the mass of the scalar field and the loop function is~\cite{Arnan:2019uhr}
\begin{equation}\label{eq:loop_c6}
    \mathcal{F}(x)=\frac{3-3x+(1+2x) \ln x}{6\,(x-1)^2}\,.
\end{equation}
When $m_{\phi}\ll m_{\chi}\approx \Lambda$, $\O^{bsV}_6$ becomes the dominant operator emerging from the $\phi \chi$ loop, thanks to the logarithmic enhancement in Eq.~(\ref{eq:loop_c6}). The phenomenology of this ``split'' dark sector was analysed  in Ref.~\cite{Darme:2020hpo} and $q^2$-dependent operators were considered also in Refs.~\cite{Datta:2017ezo,Datta:2017pfz,Datta:2018xty}. 

Analogous to this is the case where fermion and scalar multiplet charges are swapped in Eq.~(\ref{eq:np_fields})~\cite{Darme:2020hpo}, and UV constructions along similar lines 
can be considered to generate the DEFT operators involving the muon current in Eqs.~(\ref{eq:mu_oper_4}), (\ref{eq:mu_oper_6}).

Note that the DEFT coefficients $\C^{bsV}_4$, $\C^{bsV}_6$
-- similarly to $C^{\mu}_{9(10)}$ in the WET -- do not run at the leading order in QCD, since their colour part is simply a vector current~\cite{Gracey:2000am}.\footnote{Conversely, a DEFT coefficient corresponding to a tensor operator 
of the type $\O^{bsV}_5=\left(\bar{s}\sigma_{\rho \sigma}P_R b\right)V^{\rho\sigma}$, 
which can also be generated at the loop level by the UV completion, has to be renormalised before being confronted with the low-energy constraints. We leave the treatment of this and other less straightforward cases to future work.}

\subsection{Loop-level}

In the case of FIPs as dark matter particles one typically obtains the Wilson coefficients of the WET from ``candy'' diagrams constructed out of the operators in Eqs.~(\ref{eq:dm_oper_1}), (\ref{eq:dm_oper_2}). By making use of a simple cut-off regularisation, one can derive at one loop the Wilson coefficient $C_9^{\mu}$ from a fermion FIP insertion, 
\begin{equation}\label{eq:c9_ferm}
C_9^{\mu\textrm{(ferm.)}} = - \frac{\mathcal{N}\, \mathcal{C}_6^{bs\chi \chi} \mathcal{C}_6^{\mu \mu \chi \chi} }{2\, \pi\, \alpha_{\textrm{em}}\,\Lambda^2}  \,.
\end{equation}
A scalar FIP leads to
\begin{equation}\label{eq:c9_scal}
C_{9}^{\mu\textrm{(scal.)}} = - \frac{\mathcal{N}\, \mathcal{C}_6^{b s S S } \mathcal{C}_6^{\mu \mu S S} }{16\, \pi\, \alpha_{\textrm{em}}\,\Lambda^2}  \,.
\end{equation}

Note that the only other scale that enters directly the WET coefficients in Eqs.~(\ref{eq:c9_ferm}) and (\ref{eq:c9_scal}), 
apart from the EWSB scale, is the $\Lambda^2$ suppression. The exact value of the coefficients will thus depend
on the specific nature of the heavy UV completion and, as a direct consequence, the light dark matter mass will have no impact on the flavour anomalies as long as its couplings to the $b-s$ and $\mu-\mu$ current are mediated by states that can be integrated out of the low-energy theory.

\section{Flavour physics constraints\label{sec:flav-phys}}

Global scans combining the constraints from the LFUV ratios $R_K$ and $R_{K^\ast}$ with the full set of $b\to s l^+ l^-$ branching ratios and angular observables convincingly show the emergence of NP effects in the WET Wilson coefficient $C_9^{\mu}$, which can be taken alone or in combinations with others. When parametrising the global fit with one NP degree of freedom, the expected size of the Wilson coefficient is $-1.1\lesssim C_9^{\mu}\lesssim -0.5$ at $2\,\sigma$ in the single-operator case, or $-0.6 \lesssim C_9^{\mu}=-C_{10}^{\mu}\lesssim -0.3$ in a linear combination of $\mathcal{O}_9^{\mu}$ and $\mathcal{O}_{10}^{\mu}$~\cite{DAmico:2017mtc,Ciuchini:2019usw,Alguero:2019ptt,Alok:2019ufo,Datta:2019zca,Aebischer:2019mlg,Kowalska:2019ley,Ciuchini:2020gvn,Hurth:2020ehu,Alda:2020okk,Altmannshofer:2021qrr,Geng:2021nhg,Alguero:2021anc,Ciuchini:2021smi}.

As we have shown in Sec.~\ref{sec:tree-mat}, the insertion of a spin-0 FIP cannot generate the WET operator $\mathcal{O}_9^{\mu}$ (or $\mathcal{O}_{10}^{\mu}$). At the same time, Eqs.~(\ref{eq:c9_ferm}), (\ref{eq:c9_scal}) show that the numerical value of $C_9^{\mu}$ 
is quite insensitive to the presence of light dark matter-like spin-1/2 and/or spin-0 states with dimension 6 interactions. 
We can thus choose to limit our attention to the only case in which the presence of a light FIP has direct 
impact on the $b \to s\, l^+ l^-$ global fit: the vector FIP. 

The constraints on $C_9^{\mu}$ and other WET coefficients derived from the flavour anomalies lead in this case 
to a typical order-of-magnitude estimate for the DEFT couplings, obtained from a spin-1 FIP exchange in $b\to s\mu \mu$,
see Eqs.~(\ref{eq:match_c9}), (\ref{eq:match_c10}). One gets
\begin{align}\label{eq:oom}
\C^{bsV}_{4+\delta_1} \C^{\mu \mu V}_{4+\delta_2} \left(\frac{q}{\Lambda} \right)^{\delta_1 + \delta_2} \approx 10^{-9} \ ,
\end{align}
where $4+\delta_{1,2}$ indicates the dimension of the corresponding operator, and $q$ is the typical energy exchange captured in the experimental bin. For example, the biproduct of operators $\O_6^{bsV}\times \O_4^{\mu \mu V}$ develops $q^2$ dependence in the couplings, while this is not the case for a combination of operators of dimension~4.


The presence of a new light vector particle in the spectrum has far reaching consequence for a variety of SM processes.  As is discussed in the literature (see, e.g., Refs.~\cite{Dror:2017nsg,Dror:2017ehi,Michaels:2020fzj}), at FIP masses far below the typical energy scale $E$ of a given SM process, the phenomenology is dominated by the FIP longitudinal polarisation $V_L$ as $V_\mu \to \partial_\mu V_L / m_V$. While this mode cancels out for dimension-6 interactions like Eqs.~\eqref{eq:bs_oper_6} and~\eqref{eq:mu_oper_6}, it dominates for the dimension-4 cases, Eqs.~\eqref{eq:bs_oper_4} and~\eqref{eq:mu_oper_4}, as they do not correspond to conserved SM fermionic currents. More precisely, the interactions presented in Eq.~\eqref{eq:bs_oper_4} and Eq.~\eqref{eq:mu_oper_4} lead to tree-level flavour violation, weak-isospin violation (since no coupling to neutrinos where included) and an axial-coupling interaction to the SM fermions. We examine in this section the most relevant of the corresponding limits.

\subsection{$B$-meson constraints}\label{sec:BtoK}

We first consider the constraints from $B$-meson physics.

\paragraph{$B_s \to \mu \mu$} Based on a physical process related to the one generating $B \to K^{(\ast)}\mu^+\mu^-$ transitions, $\textrm{BR}(B_s \to \mu^+ \mu^-)$ can provide a strong constraint on axial muon interactions. After performing a statistical combination of the full LHCb Run~2 data with the global average of Ref.~\cite{ATLAS-CONF-2020-049}, Ref.~\cite{Altmannshofer:2021qrr} finds
\begin{equation}\label{eq:bsmu_exp}
    \textrm{BR}(B_s\to \mu^+\mu^-)_{\textrm{exp.~ave.}}=\left(2.93\pm 0.35 \right)\times 10^{-9}\,.
\end{equation}

The ratio $R_{B_s}$ between Eq.~(\ref{eq:bsmu_exp}) and the SM prediction, $\textrm{BR}(B_s\to \mu^+\mu^-)_{\textrm{SM}}=\left(3.67\pm 0.15 \right)\times 10^{-9}$~\cite{Beneke:2019slt}, can be parametrised in terms of the DEFT.
For a vector FIP, exchanged between the currents $\O_4^{bsV}$ and $\Ot_4^{\mu\mu V}$, we replace the WET Wilson coefficients  
$C_{10}^{\mu}$, $C_P^{\mu}$, $C_P^{\mu\prime}$ like in  Eqs.~(\ref{eq:match_c10})-(\ref{eq:match_cpp}). Substituting into standard expressions~\cite{Bobeth:2001sq} one gets 
\begin{align}\label{eq:RBs44}
R_{B_s}  &= \frac{ \textrm{BR}(B_s\to \mu^+\mu^-)_{\textrm{exp.}} }{  \textrm{BR}(B_s\to \mu^+\mu^-)_{\textrm{SM}}} - 1 \nonumber \\
& \simeq - \frac{8\pi\mathcal{N} \, \C_4^{bsV} \Ct_4^{\mu\mu V} }{\alpha_{\textrm{em}} C_{10}^{\rm SM} \, m_V^2}  \frac{(m_V^2-\mBs^2)^2}{(m_V^2-\mBs^2)^2 + \mBs^2 \Gamma_V^2} \,,
\end{align}
where $C_{10}^{\rm SM} = -4.31$ and we have taken the small coupling limit. Equation~(\ref{eq:bsmu_exp}) excludes the SM prediction at $2\sigma$, we thus choose to impose the bound at $3\, \sigma$ to avoid cutting out all the points with a zero or extremely small NP contribution. On the other hand, we observe that Eq.~(\ref{eq:RBs44}) is negative if  $\C_4^{bsV} \Ct_4^{\mu\mu V} > 0$ , which implies that NP contributions can potentially bring 
$R_{B_s}$ closer to the measured value. As we shall see in the next section, this sign choice is indeed preferred for $m_V$ larger than a few GeV. Conversely, we shall see that 
in the low mass regime  ($m_V\lesssim 1\,\textrm{GeV}$) one gets $\C_4^{bsV} \Ct_4^{\mu\mu V} < 0$, which produces a strong upper bound at $3 \sigma $:
\begin{align}
\left| \C_4^{bsV} \Ct_4^{\mu\mu V} \right| \lesssim 1.5 \cdot 10^{-10} \left( \frac{m_V}{1 \, \textrm{GeV}} \right)^2\,.
\end{align}

Fundamentally different are the cases of the pairs of operators $\O_6^{bsV}, \Ot_4^{\mu\mu V}$ and $\O_6^{bsV}, \Ot_6^{\mu\mu V}$ for which we find that the NP contribution to $B_s \to \mu \mu$ vanishes exactly, $R_{B_s}  = 0$, since the term contributing to the axial component is cancelled by the one feeding into the pseudoscalar current.
Interestingly, this implies that increasing the sensitivity of the measurement of $\textrm{BR}(B_s\to \mu^+\mu^-)$ may provide a handle to distinguish scenarios characterised by operators of different dimension.

\paragraph{$B_s$-mixing} $B_s - \bar{B}_s$ transitions can constrain the operators 
introduced in Sec.~\ref{sec:operators}. We use~\cite{Zyla:2020zbs} 
\begin{equation}\label{eq:bsbound}
    R_{\Delta M_s}=\frac{\Delta M_s^{\textrm{exp.}}}{\Delta M_s^{\textrm{SM}}}-1=-0.09\pm 0.08
\end{equation}
to impose a bound on the DEFT Wilson coefficients of the $b-s$ current via the dictionary between the DEFT and the WET shown in Eqs.~(\ref{eq:bsmix1})-(\ref{eq:bsmix4}) of Appendix~\ref{app:A3}.  

Similarly to the $B_s\to \mu \mu$ case, the severity of the bound strongly depends on which DEFT operator is generated in the UV completion yielding the $b-s$ coupling. The operator $\O_4^{bsV}$ leads to an $m_V^2$-rescaled modification of the WET Wilson coefficients 
$C_2$, $\widetilde{C}_2$, and $C_4$.
We obtain, in the small coupling approximation,
\begin{equation}\label{eq:bsmix}
 R_{\Delta M_s}  \simeq - \frac{\, (\C_4^{bsV})^2  }{C_{1}^{\rm SM}(\mu_b) \, m_V^2}  \frac{(\mBs^2-m_V^2)(m_V^2+R_2 M_b^2)}{(m_V^2-\mBs^2)^2 + \mBs^2 \Gamma_V^2} \,,
\end{equation}
where $C_{1}^{\rm SM}(\mu_b) = 7.2\times 10^{-11}\,\textrm{GeV}^{-2}$ and $R_2(\mu_b) \simeq -0.96$ is the ratio of the matrix elements $\langle \bar{B}_s| \mathcal{O}_2(\mu_b)| B_s \rangle/\langle \bar{B}_s| \mathcal{O}_1(\mu_b)| B_s \rangle$ (see, e.g., Ref.~\cite{Aoki:2019cca} for an updated value), all computed at the $b$-quark mass scale, $\mu_b$. 
As was mentioned above, we have not run the SM Wilson coefficients since we compare operators directly at the $B_s$ scale. By combining Eq.~(\ref{eq:bsmix}) and Eq.~(\ref{eq:bsbound}) one gets a strong constraint in the $m_V\ll 1\,\textrm{GeV}$ range:  $|\C^{bsV}_4|\lesssim 2.5 \cdot 10^{-6} \, m_V/\textrm{GeV}$. 
Conversely, with UV interactions giving rise to the 
operator $\O_6^{bsV}$,  Eq.~(\ref{eq:bsbound}) becomes
\begin{equation}\label{eq:bsmixC6}
  R_{\Delta M_s}  \simeq -  \frac{\, \mBs^2 (\C_6^{bsV})^2  }{\Lambda^4  \, C_{1}^{\rm SM}(\mu_b) }  \frac{(\mBs^2-m_V^2)(\mBs^2+R_2 M_b^2)}{(m_V^2-\mBs^2)^2 + \mBs^2 \Gamma_V^2} \,.
\end{equation}
One obtains $|\C^{bsV}_6|\, M_{B_s}^2/\Lambda^2 \lesssim 5 \times 10^{-5}$ in the low $m_V$ region.

\paragraph{$B \to K^{(*)} X $} Strong limits arise from the direct measurement of the branching ratios of the $V$, both in the case of a visible resonance and that of an invisible decay. Resonant decays to visible particles are subject to extremely strong constraints from LHCb in the $m_V$ range between $2 M_\mu$ and $M_B - M_{K^*}$~\cite{Aaij:2015tna}. We apply this bound following the numerical recast described in detail in Ref.~\cite{Darme:2020hpo}.

To impose bounds on the invisible decay width we use a combination of BaBar results~\cite{delAmoSanchez:2010bk,Lees:2013kla}.\footnote{First results from a similar Belle-II search~\cite{Belle-II:2021rof} agree with Refs.~\cite{delAmoSanchez:2010bk,Lees:2013kla}.} The adopted numerical procedure is described in detail in Appendix~A of Ref.~\cite{Darme:2020hpo}.
Assuming a large $\Gamma_V$ to invisible final states in the $m_V\ll 1\,\textrm{GeV}$ range, the bin-dependent bounds on $\textrm{BR}(B \to K +\textrm{inv.})$ induce  a strong constraint on the coupling to the hadronic current. In the dimension~4 case we get $|\C_4^{bsV}| \lesssim 10^{-8}\,m_V/\textrm{GeV}$, whereas for the operator of dimension~6 we get $|\C_6^{bsV}/\Lambda^2| \lesssim 5 \cdot 10^{-9}\,\textrm{GeV}^{-2}$. 

\subsection{Lepton sector constraints \label{sec:other}}

The second set of constraints relies instead on probing the FIP interactions with muons, the most relevant of which arise from precision measurements of the $W$ and $Z$ decays.

\paragraph{$W$ and $Z$ decays}

In order to further probe the parameter space, we perform a simple recast of the result from the ATLAS~Collaboration~\cite{ATLAS:2021kog} on $pp \to \ell \ell \ell \ell $ (referred to as the $Z\to4\mu$ search hereafter) in the $Z$-boson mass window.  We implement our effective Lagrangian via \fr/UFO~\cite{Christensen:2009jx,Degrande:2011ua,Alloul:2013bka} files, then generate hadron-level events $pp \to \mu \mu \mu \mu$ within the \amc\ platform~\cite{Alwall:2014hca}, including the selection cuts from Ref.~\cite{ATLAS:2021kog}.\footnote{We include the following cuts: four-leptons invariant mass in the range $60-100$~GeV, opposite-sign dilepton pair invariant mass larger than $5$~GeV, $p_T > 20\, \rm GeV$ for the leading lepton and $p_T > 10\, \rm GeV$ for the sub-leading lepton, and an angular separation cut as detailed in Ref.~\cite{ATLAS:2021kog}.} Our simple simulation chain leads to a SM fiducial cross-section of $22 \, \rm fb$, in good agreement with the predicted SM result from Ref.~\cite{ATLAS:2021kog}, $21.2 \pm 1.3 \, \rm fb$. We therefore use the final measured result of $22.1 \pm 1.3 \, \rm fb$ to constrain our parameter space. Note that interference with the SM plays an important role in the final cross-section computation.\footnote{A somehow similar search was performed by the CMS Collaboration~\cite{Sirunyan:2018nnz} with focuses on $L_\mu - L_\tau$ models. It yields constraints similar to the included ATLAS analysis.}

When the FIP is produced on-shell, an additional constraint on its couplings to muons arises from precision measurements of Drell-Yan dimuon production, as was shown in Ref.~\cite{Bishara:2017pje}. We find, however, that this limit is of the same order as the  $Z\to 4\mu$ ATLAS bound, or even subdominant, for the mass range $m_V=1-5\,\textrm{GeV}$ in which it was quoted. Note that the on-shell $V$ production, $Z \to \mu \mu V$, participates directly in the $Z\to 4\mu$ search and introduces a dependence of this limit to the invisible decay width of $V$.

In the low $m_V$ regime, the presence of the massive vector FIP longitudinal mode leads to a $E^2/m_V^2$ enhancement of various SM decay widths. In particular, in the limit $M_\mu, m_V \ll M_Z,M_W$ we find
\begin{equation}
    \Gamma (W \to \nu \mu V)  \simeq  \frac{\left(\C_4^{\mu\mu V}-\Ct_4^{\mu\mu V}\right)^2 G_F M_W^5}{512 \sqrt{2} \pi ^3 m_V^2}\,,\label{eq:Wwidth}
 \end{equation}    
    and
    \begin{equation}
         \frac{ \Gamma (W \to \nu \mu ) +  \Gamma (W \to \nu \mu V)}{\Gamma (W \to \nu e)}  \simeq 1 + \frac{\Gamma (W \to \nu \mu V)}{\textrm{BR}_{W \to e \nu} \Gamma_W}\,. \label{eq:sumwidth} 
    \end{equation}
We then use the world experimental average on the ratio $ \Gamma (W \to \nu \mu ) / \Gamma (W \to \nu e )$,
$0.996 \pm 0.008$~\cite{Zyla:2020zbs}, to derive the $2\,\sigma$ limit 
\begin{align}
    \left| \C_4^{\mu\mu V} - \Ct_4^{\mu\mu V}\right| \leq 0.004\left( \frac{m_V}{100 \, \rm MeV}\right)\,\label{eq:LFVWdecay} \ ,
\end{align}
which holds as long as $V$ decays mostly invisibly.
A more conservative limit (which would also apply in presence of electron couplings) is simply to require Eq.~(\ref{eq:Wwidth}) to be smaller than the total uncertainty on the measured $W$ width. Using $\Gamma_W = 2.085\pm 0.042 \, \rm GeV$ from Ref.~\cite{Zyla:2020zbs}, one obtains the $2\,\sigma$ bound
\begin{align}
    \left| \C_4^{\mu\mu V} - \Ct_4^{\mu\mu V}\right| < 0.022 \left( \frac{m_V}{100 \, \rm MeV}\right)\,,\label{eq:like65}
\end{align}
which agrees with Ref.~\cite{Karshenboim:2014tka}.

\paragraph{Resonance search in B-factories}
The BaBar~Collaboration has searched for the process $e^+ e^- \to \mu^+ \mu^- V$, $V \to \mu^+ \mu^-$, when the FIP is assumed to have a small width and a mass up to around $10\,\textrm{GeV}$~\cite{TheBABAR:2016rlg}. In the relevant mass region, our model requires in any case a large invisible width to escape resonant $B \to K^{*} \mu \mu$ searches, so that this constraint is subdominant. 
The Belle-II Collaboration recently provided a bound 
on the final-state radiation process $e^+ e^- \to \mu^+ \mu\, V$, 
$V \to \textrm{invisible}$, based on $0.28\,\textrm{fb}^{-1}$ of data from the 2018 run~\cite{Belle-II:2019qfb}. 
While the current limit can hardly compete with other bounds
the 2019 run has stored $\sim 10\,\textrm{fb}^{-1}$ of data so that the search is expected to become rapidly relevant in the near future.

\paragraph{Coupling to electrons} We choose not to consider in this work explicit interactions of the FIP with electrons. Nevertheless, a coupling to electrons is generated via the vector kinetic mixing when at least one of the DEFT operators is at dimension~4~\cite{Holdom:1985ag}. Thus, even below the di-muon threshold, we expect $V$ to decay visibly into $e^+e^-$ in the absence of an invisible decay channel. Constraints on a light vector FIP coupled to electrons were discussed, e.g., in Refs.~\cite{Altmannshofer:2017bsz,Datta:2017ezo}. In particular, resonance searches at low $q^2$ in LHCb~\cite{Aaij:2013hha,Aaij:2015dea} exclude the parameter space relevant for the $b \to s$ anomalies, thus requiring the $V$ to decay mostly invisibly (in which case the constraints on $B\to K + \rm inv.$ described above apply).\footnote{LHCb low-resonance searches extend down to $\sim 20\,\rm MeV$. 
A very light electron-philic FIP is therefore not constrained by this approach. 
However, stringent limits on long-lived dark photons then apply, with the parameter space almost completely covered~\cite{Beacham:2019nyx}.}

\subsection{Muon anomalous magnetic moment \label{sec:anom_gm2}}
If the flavour anomalies are explained by a vector FIP exchange, 
the same FIP can potentially contribute to the anomalous magnetic moment of the muon. The recent confirmation at Fermilab~\cite{PhysRevLett.126.141801} 
of a $3.3\,\sigma$ discrepancy between the observed value of $(g-2)_{\mu}$ and the SM
renders this possibility all the more enticing and timely. 

The computation of $(g-2)_\mu$ can be enhanced at the 1-loop level 
if the vector FIP $V$ interacts with the muon current via $\O_4^{\mu\mu V}$ and $\Ot_4^{\mu\mu V}$. One gets~\cite{Jegerlehner:2009ry,Queiroz:2014zfa}
\begin{equation}
\label{eq:gm2}
\delta(g-2)_\mu =  \frac{1}{8 \pi^2} \frac{\mmu^2}{\mv^2} \left[(\C_4^{\mu\mu V})^2 \mathcal{F} \left(x_\mu \right) + (\Ct_4^{\mu\mu V})^2 \widetilde{\mathcal{F}} \left(x_\mu \right) \right]\,,
\end{equation}
where $x_\mu = \mmu/\mv$ and the loop functions read
\begin{align}\label{eq:gm2loop}
\mathcal{F}(x) &=\int_0^1 dz\, \frac{2 z^2 \left(1-z\right)}{x^2 z + (1-z)(1-x^2z)}\,, \\
\widetilde{\mathcal{F}}(x) &=\int_0^1 dz\, \frac{ \left[2z \left(1-z\right) \left(z-4\right) - 4x^2 z ^3\right]}{x^2 z + (1-z)(1-x^2z)}\,.
\end{align}

Conversely, no significant enhancement is obtained if the coupling of the FIP 
to the muon proceeds through the operators $\O_6^{\mu \mu V}$ and 
$\Ot_6^{\mu \mu V}$. In those cases the value of $(g-2)_\mu$ is suppressed by the UV cut-off $\Lambda$ yielding, for example,
\begin{eqnarray}
\dgmtwo & = & 
\frac{\left( \mathcal{C}_6^{\mu\mu V} \right)^2 M_{\mu}^2}{12\,\pi^2 \Lambda^2} \,,
\label{eq:gm2O66}
\end{eqnarray}
in the case of $\O_6^{\mu \mu V}$ and a similar expression for $\Ot_6^{\mu \mu V}$.
Its exact numerical value depends entirely on the specifics of the UV completion.

As we have described in Sec.~\ref{sec:BtoK}, 
the viable range of the DEFT coefficients $\C_4^{bsV}$, $\C_6^{bsV}$, 
corresponding to the operators of the $b-s$ current, is bounded by the constraints 
from $B_s$-mixing and $B\to K^{(\ast)}\,X$ searches. As a direct consequence, for some points in the numerical scan the typical values of $\C_4^{\mu\mu V}$ required for a reasonable 
agreement with the flavour anomalies is very large, leading to a deviation in $\delta(g-2)_{\mu}$ widely exceeding the measured value. A cancellation with the contribution from the axial-vector coupling $\Ct_4^{\mu\mu V}$ may therefore be necessary~\cite{Sala:2017ihs}.
In particular, as we will show in the next section, such cancellation is always required for light $m_V$  whereas it is not needed for a GeV-scale FIP.
Note that including the axial-vector contribution can trigger the bounds from $B_s \to \mu \mu$, also discussed in Sec.~\ref{sec:BtoK}, which are particularly strong in the $m_V\ll 1\,\textrm{GeV}$ range.

\section{Numerical results\label{sec:numerics}}

In the rest of this work we will consider somehow arbitrarily FIP masses up to $20\,\textrm{GeV}$, so that $m_V^2 / \Lambda^2_{\rm EW} \ll 1$. While there is no specific upper bound on the mass of the vector FIP from the constraints listed in the previous section, we also do not include in our simplified models the interactions between $V$ and the electroweak sector. We leave the complete study of the ``transitional'' regime for larger masses, up to the electroweak scale $\Lambda_{\rm EW}$, for future work.

\subsection{Fit to the \texorpdfstring{$\boldsymbol{b \to s}$}{b to s} anomalies\label{sec:fit}}

We perform a multidimensional fit to the $b \to s $ anomalies, including in particular the LFUV ratios $R_K$~\cite{Aaij:2021vac} and $R_{K^*}$~\cite{Aaij:2017vbb}, the most updated $B\to K^{\ast}\mu^+\mu^-$ angular-observable data~\cite{Aaij:2020nrf}, and the finely binned results for the $B\to K\mu^+\mu^-$ and $B\to K^{\ast}\mu^+\mu^-$ branching ratios~\cite{LHCB:2014cxe,Aaij:2016flj}. We scan over the following free parameters: $m_V$, $\gamma_V\equiv\Gamma_V/m_V$, $\C_{4,6}^{bsV}$, $\C_{4,6}^{\mu\mu V}$, and $\Ct_{4,6}^{\mu\mu V}$. The vector mass $m_V$, expressed in GeV, is flatly distributed either in the $[0.01, 2]$ range, or in the $[2, 20]$ one. For the ratio between the $V$ width and its mass, $\gamma_V$, we employ a logarithmically-flat prior in the range $[10^{-3}, 0.5]$. All the absolute values of the DEFT coefficients have a logarithmically-flat distributed prior in the range $[10^{-10}, 1]$. Since the observables included in these fits depend only on the products $\C_i^{bsV}\cdot\C_j^{\mu\mu V}$ and $\C_i^{bsV}\cdot\Ct_j^{\mu\mu V}$, but not on the single coefficients, the fit results presented in this subsection are given in terms of DEFT biproducts rather than as a function of individual coefficients.

As was described in detail in Ref.~\cite{Darme:2020hpo}, we perform separate fits depending on whether $m_V$ lies above or below $2\,\textrm{GeV}$. 
In fact, in order to obtain an adequate fit, one has to require that the product $\C_i^{bsV}\cdot\C_j^{\mu\mu V}$ $(\C_i^{bsV}\cdot\Ct_j^{\mu\mu V})$ assumes a different sign in each of these two regions. We note that this leads to a negative (positive) $C_9^\mu$ $(C_{10}^\mu)$, in agreement with the  global WET fits.
Following Ref.~\cite{Darme:2020hpo}, we refer to these distinct cases as the \textit{high-mass} fit and the \textit{low-mass} fit. 

Due to the $q^2$ dependence of Eqs.~\eqref{eq:match_c9}-\eqref{eq:match_cpp}, it is not possible to refer explicitly to the results of the 
global WET fits that can be found in the literature. One needs instead to confront the DEFT parameter space 
directly with the experimental data. In order to do so, we employ a custom version of the \texttt{HEPfit} package~\cite{deBlas:2019okz}, performing a Markov Chain Monte Carlo analysis by means of the Bayesian Analysis Toolkit (BAT)~\cite{Caldwell:2008fw}.

\begin{figure}[!h!]
\centering
\subfloat[]{%
\includegraphics[width=0.51\textwidth]{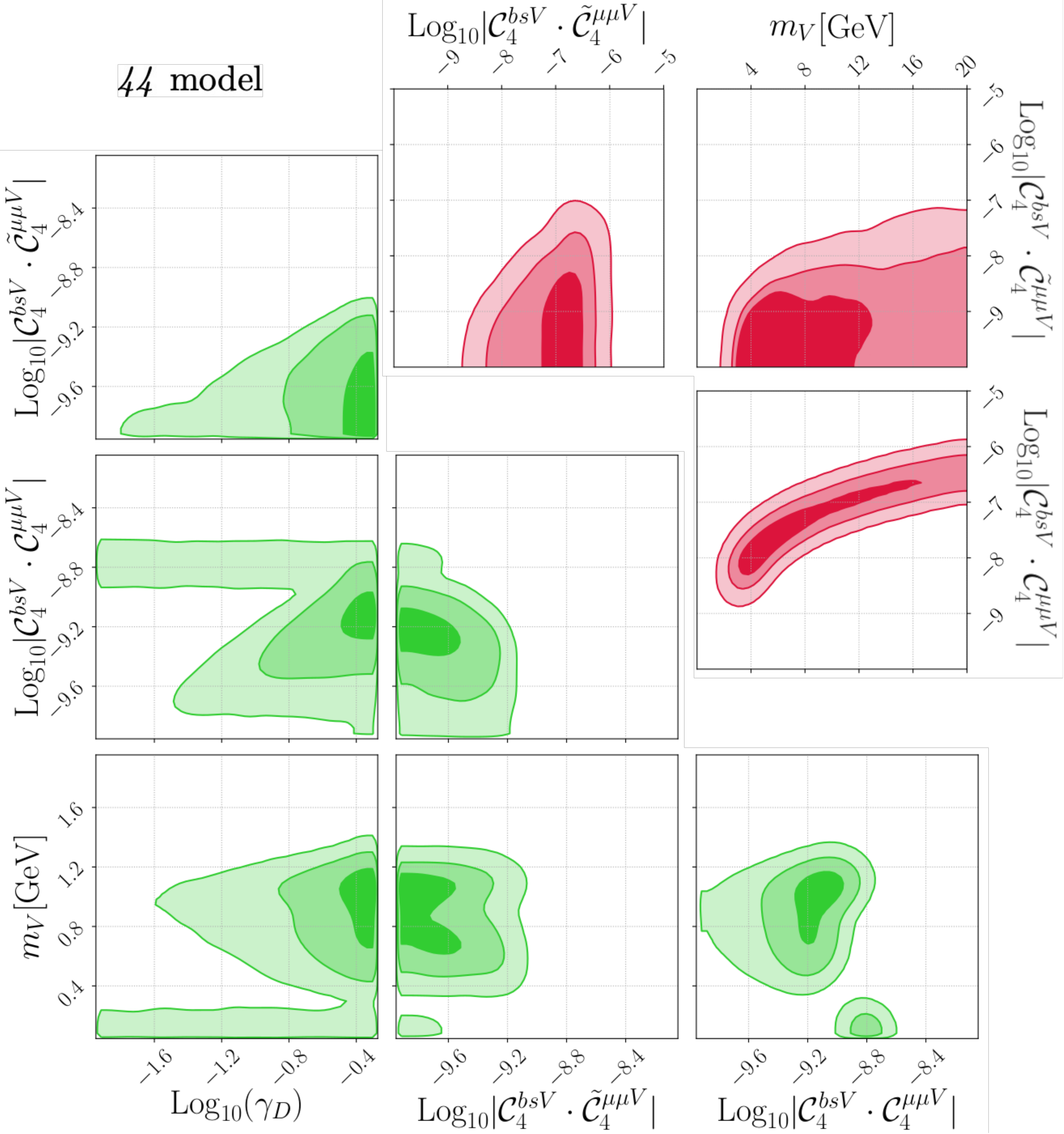}
}\\%
\subfloat[]{%
\includegraphics[width=0.51\textwidth]{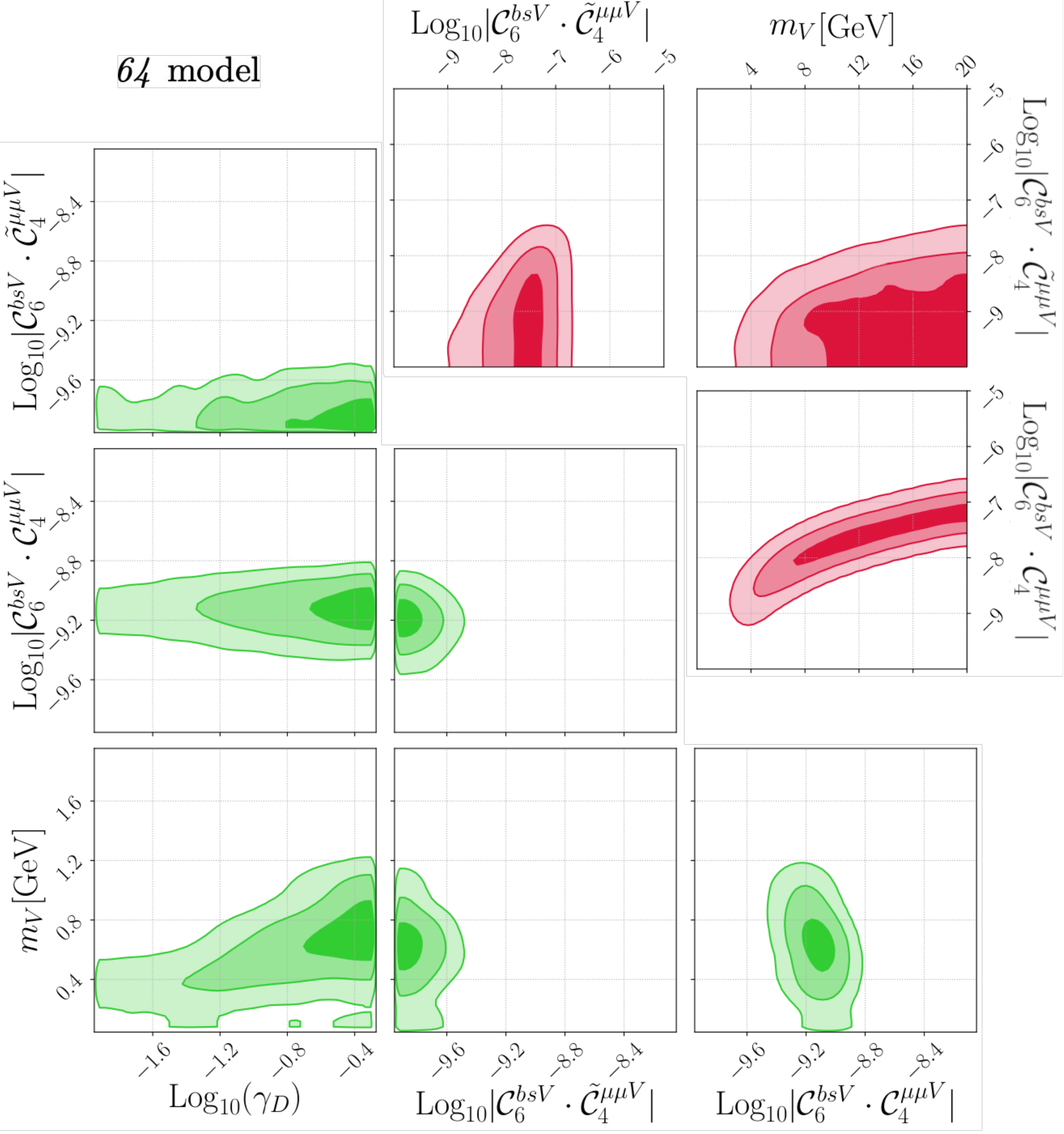}
}%
\caption{(a) Inference of model parameters from a fit to $b\to s$ anomalous data for the DEFT operators of dimension~4 in both the $b-s$ and $\mu-\mu$ currents (\textit{44}~model). (b) Same for DEFT operator of dimension~6 in the $b-s$ current and dimension 4 in $\mu-\mu$ (\textit{64}~model). 
The plots in red are relative to the high-mass fits (see main text) 
while the ones in green are relative to the low-mass ones, with the contours corresponding to the smallest regions of 68\%, 95\%, 99.7\% probability. We refer to the main text for the sign of the couplings. Note that in (b) the coupling $\C_6^{bsV}$ is reported in units of GeV$^{-2}$.
\vspace{-1cm}}
\label{fig:FitRes}
\end{figure}

We present in Fig.~\ref{fig:FitRes}(a) the results of the fit in the case of DEFT operators of dimension~4 in both the $b-s$ and $\mu-\mu$ currents (we dub this as the \textit{44~model} hereafter). The plots in red are relative to the high-mass fit, where $\C_4^{bsV}\cdot\C_4^{\mu\mu V}$ is required to be negative,
the plots in green are relative to the low-mass one, where $\C_4^{bsV}\cdot\C_4^{\mu\mu V}$ is required to be  positive,
and the contours correspond to the smallest regions of 68\%, 95\%, 99.7\% probability. In particular, 
in Fig.~\ref{fig:FitRes}(a) we are showing the marginalised 2-dimensional posterior probability density function~(2D~pdf) in planes of the vector mass $m_V$, width-to-mass ratio $\gamma_V$, and the products of DEFT coefficients $\C_4^{bsV}\cdot\C_4^{\mu\mu V}$ and $\C_4^{bsV}\cdot\Ct_4^{\mu\mu V}$. In accordance with the respective priors, the posterior pdf's for the width and the coefficients are reported in logarithmic scale. The 2D~pdf's involving $\gamma_V$ are only shown for the low-mass fit since they are flatly distributed in the high-mass case and hence not particularly informative.

In the high-mass region of Fig.~\ref{fig:FitRes}(a) (in red) one can observe a strong correlation between the vector mass $m_V$ and the coupling product $\C_4^{bsV}\cdot\C_4^{\mu\mu V}$, responsible for the NP contributions to $C_9^{\mu}$, cf.~Eq.~\eqref{eq:match_c9}. The coupling biproduct can range from $\sim 10^{-8}$ (corresponding to $m_V \approx 4$~GeV) up to $\sim 10^{-6}$ at the $2\,\sigma$ level, if $m_V$ grows as well. From the correlation between $\C_4^{bsV}\cdot\C_4^{\mu\mu V}$ and $\C_4^{bsV}\cdot\Ct_4^{\mu\mu V}$ it is also possible to observe that the latter can grow from negligible values up to the size of the former (at the $3\,\sigma$ level) but never get larger. This is consistent with the results of global WET fits~\cite{DAmico:2017mtc,Ciuchini:2019usw,Alguero:2019ptt,Alok:2019ufo,Datta:2019zca,Aebischer:2019mlg,Kowalska:2019ley,Ciuchini:2020gvn,Hurth:2020ehu,Alda:2020okk,Altmannshofer:2021qrr,Geng:2021nhg,Alguero:2021anc,Ciuchini:2021smi}, where NP contributions to $C_{10}^{\mu}$ -- which, as Eq.~\eqref{eq:match_c10} shows, in the DEFT arise precisely from $\C_4^{bsV}\cdot\Ct_4^{\mu\mu V}$ -- are, if non-vanishing, usually smaller in size than the ones to $C_9^{\mu}$.\footnote{Exceptions to this statement apply if one advocates for a very conservative treatment of the hadronic uncertainties or, alternatively, for an additional universal NP component in both the electron and muon vector currents. Indeed, if one allows for either one of these possibilities it becomes easier to accommodate NP contributions to the muon axial current of the same size as the ones in the muon vector current (and hence $|\C_4^{\mu\mu V}| \approx |\Ct_4^{\mu\mu V}|$)~\cite{Ciuchini:2019usw,Alguero:2019ptt,Aebischer:2019mlg,Ciuchini:2020gvn,Altmannshofer:2021qrr,Alguero:2021anc,Ciuchini:2021smi}.} Note that the solution identified in Ref.~\cite{Sala:2017ihs}, corresponding to a mediator with mass $m_V \approx 2.5\,\textrm{GeV}$ is disfavoured in our study. 
This is due to primarily two reasons. On the one hand, the set of observables in Ref.~\cite{Sala:2017ihs} included only the LFUV ratios and $P_5'$, 
neglecting other measurements, e.g., the branching ratio $\textrm{BR}(B\to K^{\ast}\mu^+\mu^-)$~\cite{Aaij:2016flj}, which generate large chi-squared values for $m_V\approx 2.5\,\textrm{GeV}$.
On the other hand, the recently updated values for the angular analysis of $B\to K^{\ast}\mu^+\mu^-$, and hence for $P_5'$, also push the favoured values for $m_V$ above the region highlighted in Ref.~\cite{Sala:2017ihs}.

In the low-mass region (in green) $\C_4^{bsV}\cdot\C_4^{\mu\mu V}$ ranges between $10^{-10.0}-10^{-8.6}$, in correspondence of which $m_V$ can assume two distinct values, one peaked around $1\,\textrm{GeV}$ and the other for $m_V\lesssim 200\,\textrm{MeV}$. The origin of these solutions can be traced back to the peculiar $q^2$ dependence of Eqs.~\eqref{eq:match_c9}, \eqref{eq:match_c10},
which can give rise, for specific choices of $m_V$, to destructive and/or constructive interference in the DEFT Wilson coefficients that are fitted to the experimental bins. In particular we observe that, while at $m_V\approx 1\,\textrm{GeV}$ the global fit is dominated by the branching ratio $\textrm{BR}(B\to K \mu^+\mu^-)$, for the region at lower masses the prevalent channel is given by the ``angular''~$P_5'$, which requires 
the biproduct $\C_4^{bsV}\cdot\C_4^{\mu\mu V}$ to be larger than $\C_4^{bsV}\cdot\Ct_4^{\mu\mu V}$.
Note that the solution with $m_V\approx 1\,\textrm{GeV}$ is characterised by a non-negligible width, as highlighted by the peak around $\gamma_V\approx 0.4-0.5$ in the posterior pdf. This does not seem to be the case instead for $m_V\lesssim 200\,\textrm{MeV}$, thus highlighting the prevalence of different observables in the two regions.

The fit results in the case of DEFT operators of dimension~6 in the $b-s$ current and dimension 4 in $\mu-\mu$ are presented in Fig.~\ref{fig:FitRes}(b) (we dub this as the \textit{64~model} hereafter). The colour scheme is the same as in Fig.~\ref{fig:FitRes}(a). $\C_6^{bsV}\cdot\C_4^{\mu\mu V}$ is required to be negative in the high-mass region (in red), and positive in the low-mass one (in green).
Quantitative differences with respect to the \textit{44}~model can be found in this case, driven by the different $q^2$ dependence of the NP contributions, but we can also highlight a few qualitative similarities, for example, there remains a strong correlation between $m_V$ and the coupling product $\C_6^{bsV}\cdot\C_4^{\mu\mu V}$ in the high-mass region. However, the high-probability region of the pdf is peaked now at higher $m_V$. In the low-mass region one can see that $\C_6^{bsV}\cdot\C_4^{\mu\mu V}$ is bound to a narrow range whereas $m_V$ follows a distribution similar to the \textit{44}~model. The region of high probability at $m_V\approx 1\,\textrm{GeV}$ appears more pronounced in the \textit{64}~model than in the \textit{44}~model, and it seems to require a comparatively smaller width. This is due to the explicit $q^2$ dependence of the $bsV$ coupling in the \textit{64}~model, which facilitates a good fit to many 
of the $\textrm{BR}(B\to K \mu^+\mu^-)$ data bins.

Note that a fit performed in the case of DEFT operators of dimension~4 in the $b-s$ current and dimension 6 in $\mu-\mu$ (\textit{46~model}) yields results perfectly analogous to the ones shown in Fig.~\ref{fig:FitRes}(b), given the equal $q^2$ dependence of the two scenarios, presented in Eqs.~\eqref{eq:match_c9}, \eqref{eq:match_c10}. 

\begin{figure}[!t!]
\begin{center}
\includegraphics[width=0.52\linewidth]{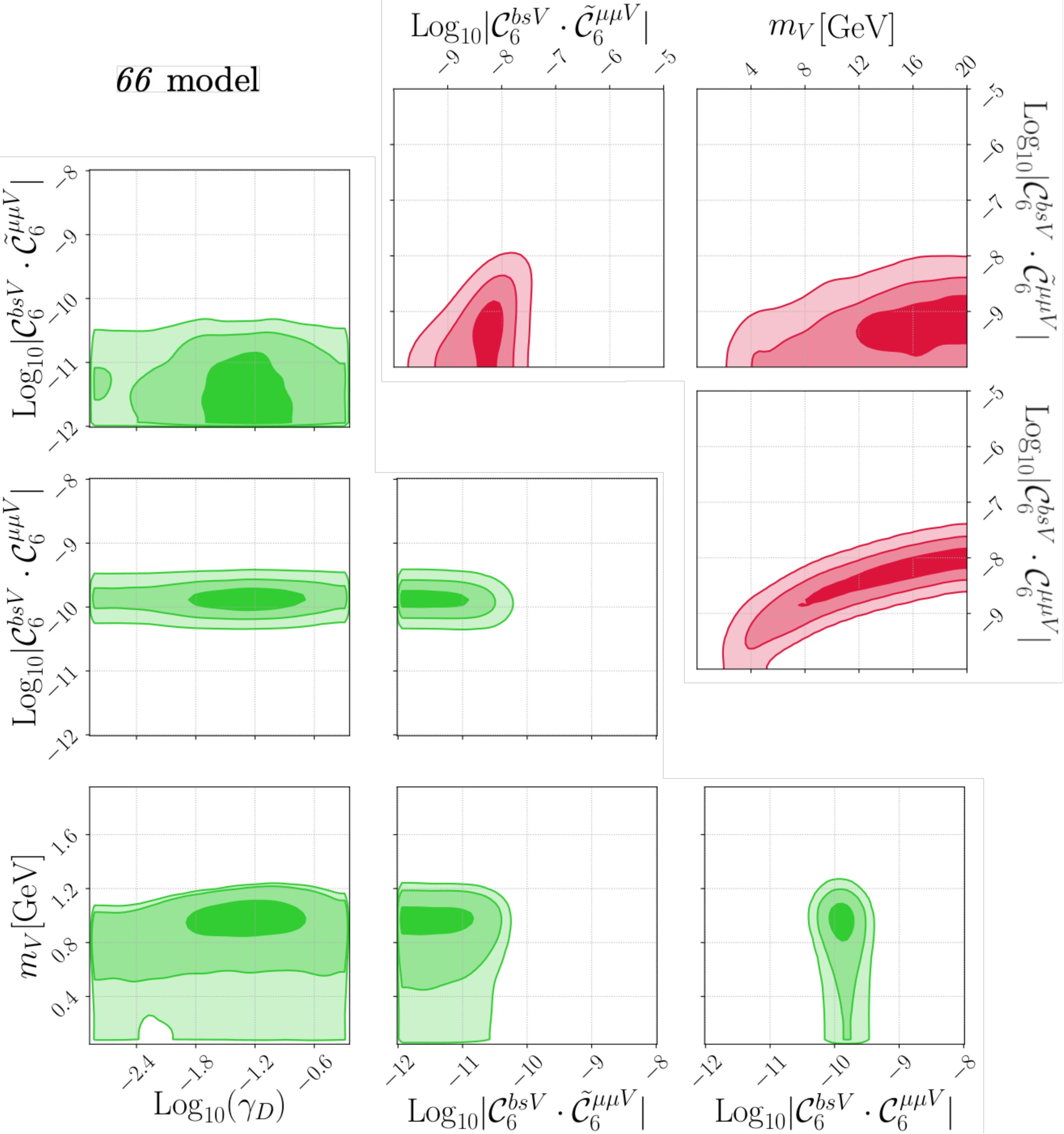}
\caption{Inference of model parameters from a fit to $b\to s$ anomalous data with DEFT operators of dimension~6 in both the $b-s$ and $\mu-\mu$ currents (\textit{66}~model). The colour scheme is the same as in Fig.~\ref{fig:FitRes}. Note that the couplings $\C_6^{bsV}$, $\C_6^{\mu\mu V}$ and $\Ct_6^{\mu\mu V}$ are reported in units of GeV$^{-2}$.}
\label{fig:FitRes2}
\end{center}
\end{figure}

Finally, we present in Fig.~\ref{fig:FitRes2} the marginalised 2D~pdf of the fit in the case of DEFT operators of dimension~6 in both the $b-s$ and $\mu-\mu$ currents (\textit{66~model}). The colour scheme is the same as in Fig.~\ref{fig:FitRes}. The fit shows many qualitative similarities to the \textit{64} case, with notable differences pertaining mostly to the favoured values of the couplings, due to the different $q^2$ dependence, and a pronounced high-probability peak at $m_V\approx 1\,\textrm{GeV}$, $\gamma_V\approx 10^{-1.8}-10^{-1}$, with lower probability when the $V$ mass becomes smaller. Note how the transition to the region where the angular observables exert the greatest relative pull, at very low $m_V$, is much smoother in this model than in the previous cases, indicating that it is not difficult to fit the $\textrm{BR}(B\to K \mu^+\mu^-)$ bin data over a broader mass range, due the additional $q^2$-dependence of the muon coupling.

\subsection{Including all flavour constraints \label{sec:cons_app}}

\begin{figure}[t]
\centering
\subfloat[]{%
\includegraphics[width=0.65\textwidth]{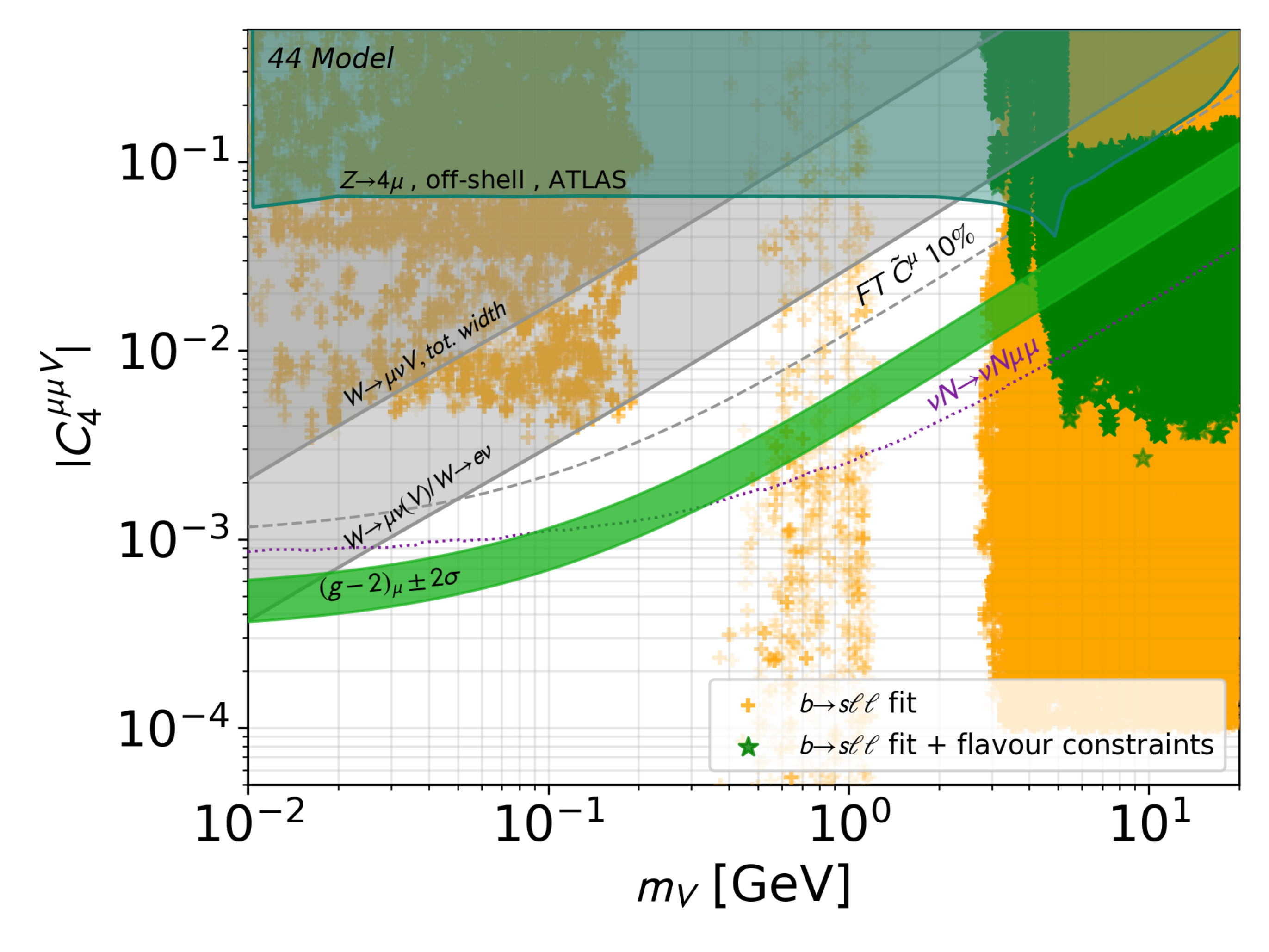}
}%
\hspace{0.02\textwidth}
\subfloat[]{%
\includegraphics[width=0.65\textwidth]{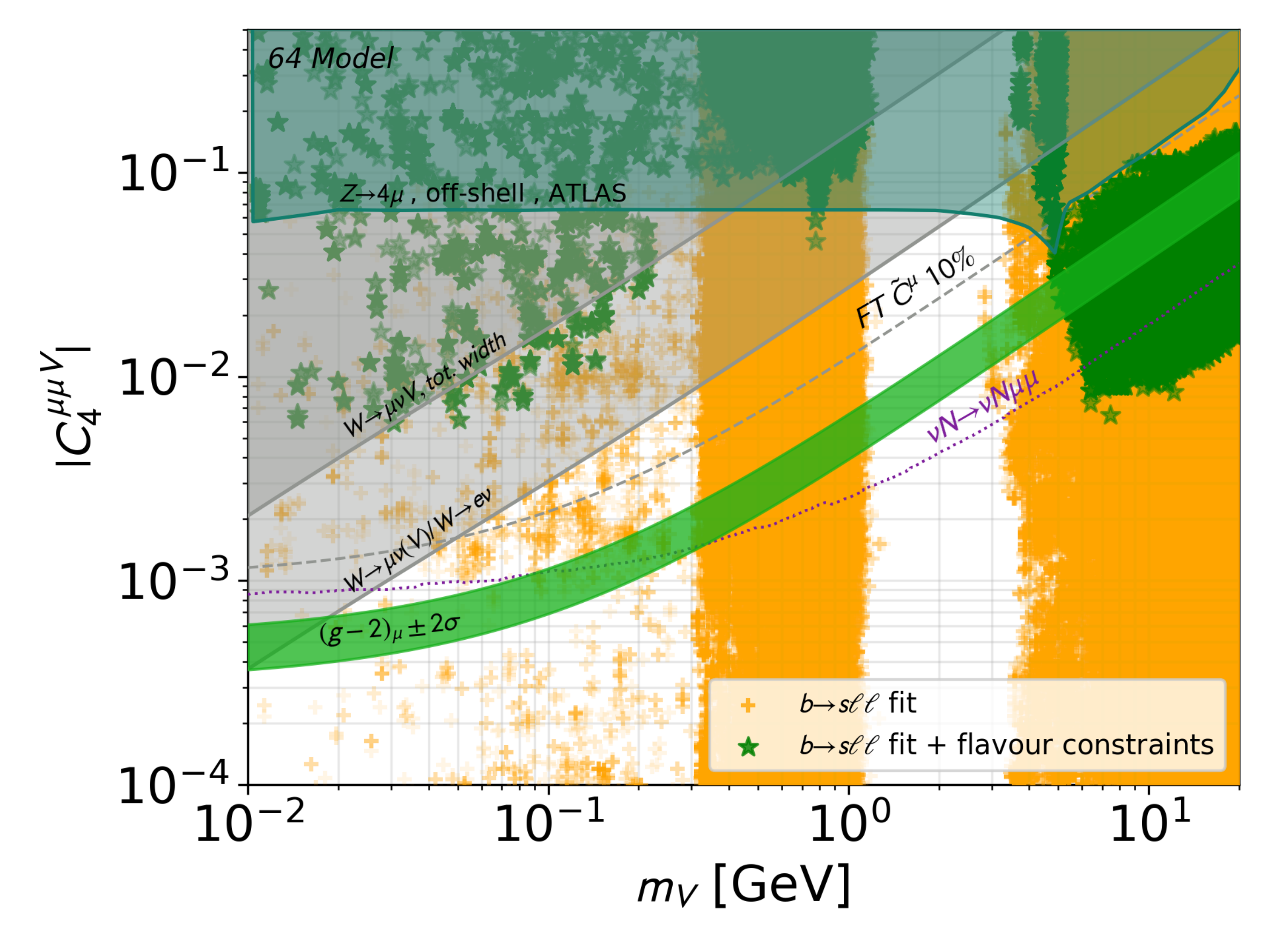}
}%
\caption{Constraints on the parameter space consistent with the LFUV anomalies (yellow points) 
in the plane of $\C_4^{\mu\mu V}$ versus the vector mass $m_V$. 
(a) The \textit{44}~model. (b) The \textit{64}~model. Besides being consistent at $2\,\sigma$ with the $b \to s$ anomalies, all green points satisfy the constraints from $B\to K +\rm inv.$, $B \to K^* \mu \mu$, $B_s-\bar{B}_s$ mixing, $B_s \to \mu \mu$ and $Z\to 4\mu$ (off-shell+on-shell) at $2\sigma$, applied following the procedure described in the text. We show overlaid the bounds that do not depend on the hadronic sector and/or on $\Gamma_{V,\textrm{inv}}$. The blue region covers the parameter space excluded by the multilepton ATLAS search~\cite{ATLAS:2021kog} (off-shell), whereas the grey regions show bounds from LFUV in $W$-decay (see main text). The green band is consistent with the $(g-2)_\mu$ measurement~\cite{PhysRevLett.126.141801} at $2\,\sigma$, with no fine tuning of the couplings. All points above the green band require a non-zero $\Ct_4^{\mu\mu V}$, with the dashed grey line indicating a $10 \%$ fine tuning with $\C_4^{\mu\mu V}$.
The dotted purple line in both panels shows the upper bound from neutrino trident production~\cite{Altmannshofer:2014pba}, 
obtained under the assumption that the coupling of $V$ to neutrinos is of the same size as $\C^{\mu\mu V}_4$.
\vspace{-2cm}}
\label{fig:chainC44}
\end{figure}

We can now apply the constraints introduced in Secs.~\ref{sec:BtoK}, \ref{sec:other} to the high-probability regions of the fits described above. We present the results for the $\textit{44}$~model in Fig.~\ref{fig:chainC44}(a), in the plane of the vector muon coupling versus the FIP mass $m_V$. The yellow points in the figure are within the $2\,\sigma$ regions of the fits to $b \to s $ anomalies
described in Sec.~\ref{sec:fit}. Green points satisfy, additionally,  
$B_s-\bar{B}_s$ mixing and $B_s \to \mu \mu$ constraints, as well as ATLAS $Z \to\mu \mu V^{(\ast)}$ at the $2\,\sigma$ level, as described in Sec.~\ref{sec:other}.
We have also ensured that all of them pass the constraints on invisible final states from the BaBar searches~\cite{delAmoSanchez:2010bk,Lees:2013kla}, numerically recast as described in Ref.~\cite{Darme:2020hpo}, and the bounds from resonant $B\to K^{\ast}\mu^+\mu^-$
at LHCb, also described in Ref.~\cite{Darme:2020hpo}. We finally overlay
the remaining limits, which are independent of $\Gamma_{V,\textrm{inv}}$. The green band corresponds to $\C_4^{\mu\mu V}$ being consistent at $2\,\sigma$ with the $(g-2)_\mu$ measurement~\cite{PhysRevLett.126.141801}.  

Some of the constraints 
applied to the parameter space depend explicitly on both $\C_4^{\mu\mu V}$ and $\Ct_4^{\mu\mu V}$. This is the case, particularly, of the $W$-decay bound of Eq.~(\ref{eq:Wwidth}) and forward, and the numerical recast of the $Z\to 4\mu$ cross-section bound. When these bounds are applied to regions of the parameter space for which 
$\C_4^{\mu\mu V}$ is too large to yield a value of $\dgmtwo$ in agreement with the experimental determination, we tune $\Ct_4^{\mu\mu V}$ as required to bring $\dgmtwo$ down to the measured $2\,\sigma$ region, cf.~Sec.~\ref{sec:anom_gm2}.\footnote{As the sign of $\Ct_4^{\mu\mu V}$ is not fixed by this procedure, we choose it in agreement with the requirements from the $b \to s$ fits, i.e.,  opposite to the sign of $\C_4^{\mu\mu V}$.} The dashed grey line in the figure indicates a fine tuning at the 10\% level.

FIPs with mass above the $B$-meson can easily yield an excellent fit to the experimental data while escaping all current constraints. This conclusion holds for the \textit{44}~model, shown in Fig.~\ref{fig:chainC44}(a), as well as for the \textit{64}~model, shown in Fig.~\ref{fig:chainC44}(b). The favoured parameter space in the $\textit{64}$~model is in agreement with the results obtained in the UV model of Ref.~\cite{Darme:2020hpo}.  Incidentally, we find it remarkable that these very compact simplified-model solutions to the $b \to s $ anomalies feature also excellent prospect to solve the $(g-2)_\mu$ anomaly with no fine tuning required. In the high mass region, the lower limit on the green points distribution arises from the $B_s$-mixing constraints on $\C_4^{bs V}$ and $\C_6^{bs V}$ and the upper limit from the ATLAS~Collaboration~\cite{ATLAS:2021kog} search for $Z\to4\mu$.

We observe in Fig.~\ref{fig:chainC44}(b) that for the \textit{64}~model the fit yields a large number of points compatible with the various $B$-physics constraints in the very low-mass region of the parameter space, $m_V\ll 2\,M_{\mu}$. On the other hand, the required couplings to the muon are typically quite large, to overcome the strong bounds on $\C_6^{bsV}/\Lambda^2$ from $B\to K + \textrm{invisible}$. This leads to $\dgmtwo$ exceeding the experimental value and, consequently, a unified solution to all anomalies requires the aforementioned 
fine tuning of the axial-vector muon coupling against the vector one.
The overwhelming majority of these points are excluded by flavour-universality tests in $W$ decays, as measured at the LHC~\cite{Zyla:2020zbs}. 

\begin{figure}[!t!]
\begin{center}
\includegraphics[width=0.65\linewidth]{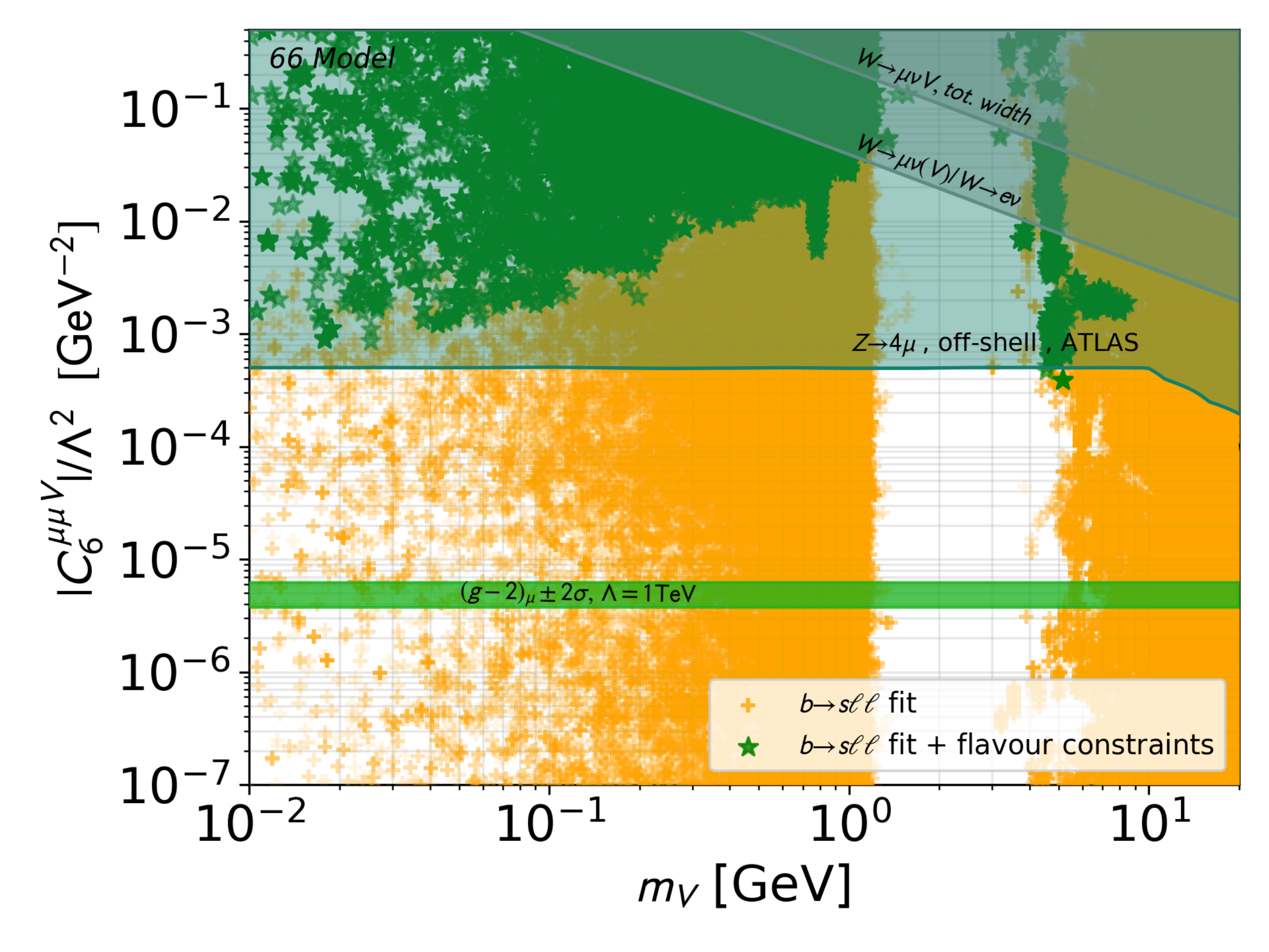}
\hspace{0.cm}
\caption{Constraints on the parameter space consistent with the LFUV anomalies (yellow points) 
in the plane of $\C_6^{\mu\mu V}$ versus the vector mass $m_V$ for the \textit{66}~model. The colour scheme is the same as in Fig.~\ref{fig:chainC44}. The position of the $(g-2)_\mu$ $2\,\sigma$ band corresponds to a 
UV scale $\Lambda = 1\, \rm TeV$, see Eq.~\eqref{eq:gm2O66}.
\label{fig:chainC66}}
\end{center}
\end{figure}

We show in Fig.~\ref{fig:chainC66} the corresponding results for the \textit{66}~model. There appear to be extremely few points that can pass all the cuts simultaneously, with the limit from the multilepton ATLAS search~\cite{ATLAS:2021kog} dominating the constraints on the muon coupling. This exclusion reflects the fact that the FIP coupling to the muon, $\C_6^{\mu \mu V}$, is of dimension~$6$ and thus more sensitive to high-energy processes than in the \textit{44} and \textit{64} cases.

We point out, finally, that in the presence of a neutrino coupling, stringent constraints from neutrino trident production apply~\cite{Altmannshofer:2014pba}, potentially excluding a common solution for the $(g-2)_{\mu}$ and LFUV anomalies in the high-mass region. For indicative purposes we show as a dotted purple line in both panels of Fig.~\ref{fig:chainC44} the upper bound obtained under the assumption that the coupling of $V$ to neutrinos is of the same size as $\C^{\mu\mu V}_4$. This would be indeed the case for SU(2)$_L$-conserving interactions of the $V$ with the lepton doublet, like in the well-known $L_\mu-L_\tau$ gauge group. However, it is possible to envision more elaborate UV constructions that could allow one to suppress the coupling to the neutrino with respect to the corresponding charged lepton, in which case the purple line in Fig.~\ref{fig:chainC44} would shift upwards.\footnote{One may construct for instance a 2-Higgs doublet model where the additional Higgs doublet and extra VL heavy leptons all carry U(1)$_D$ charge: $\Theta:(\mathbf{1},\mathbf{2},1/2,Q_D)$, $E:(\mathbf{1},\mathbf{1},1,Q_D)$. Yukawa couplings with the SM left-handed muon doublet, $L_{\mu}$, of the type $\lambda E \Theta^{\dag} L_{\mu}$, generate a left-chiral coupling 
$g_L^{\mu\mu}$ of $V$ to the muons once U(1)$_D$ is broken. The coupling to neutrinos is absent at the tree level. Typically one gets
$g_L^{\mu\mu}\approx -g_D Q_D \lambda^2 v_D^2/2 m_E^2$. 
The spectrum will have to feature additional scalar singlets/multiplets of SU(2)$_L$, 
whose scalar potential and dark charges have to be adjusted to cancel the direct $V/Z$ mixing and to raise the mass of the charged scalars above direct LHC bounds.} Alternatively, the presence of extra particles and interactions in the UV may introduce contributions to the $(g-2)_{\mu}$ beyond the ones computed in our bottom-up approach. The green band in Fig.~\ref{fig:chainC44} would in that case shift downwards.

\section{Summary and conclusions \label{sec:summary}}

In this paper we have performed a comprehensive phenomenological analysis of a set of simplified models providing a 
combined solution to the anomalous magnetic moment of the muon and 
the flavour anomalies emerged at LHCb in $b\to s l^+ l ^-$ transitions (both flavour-conserving and lepton-flavour violating). Our simplified models are based on a set of operators of the Dark EFT, or DEFT, where with this term we encompass a broad range of interactions between the SM and one or more light, feebly interacting particles (FIPs) that find their origin in heavy degrees of freedom integrated out of the theory. The most promising models for fitting both anomalies are based on vector FIPs, with the FIP exchange at low energy giving rise to the anomalies in flavour and $(g-2)_{\mu}$.

We have divided our simplified models in categories determined by the explicit momentum dependence 
of the FIP interaction with the $b-s$ and $\mu-\mu$ vector currents. For all categories, after performing a global fit to the experimental data we have applied a large set of constraints to the favoured parameter space, extracted from direct measurements of the $B\to K +\textrm{invisible}$ branching ratio, $B\to K^{\ast}\mu^+\mu^-$ resonant decays, the measurements of 
$\textrm{BR}(B_s\to \mu^+\mu^-)$, and $B_s$-mixing. Moreover, we have computed for the first time the strong 
constraints arising from several precision measurements associated
with $W$- and $Z$-boson decay, namely the measurement of the $Z\to 4 \mu$ cross section at ATLAS and the decay 
$W\to \mu+\textrm{invisible}$. We find that unified explanations of the $(g-2)_{\mu}$ and $b\to s l^+ l ^-$ anomalies exist and pass all the constraints in a model based on vector-mediator exchange with DEFT operators of dimension~4 in both the $b-s$ and $\mu-\mu$ vector currents (model~\textit{44}, with no momentum dependence of the couplings), and in vector-mediator exchange with dimension~6 in the $b-s$ current ($q^2$-dependence of the coupling) and dimension~4 in $\mu-\mu$ (we call it model~\textit{64}). 
In both models the typical range of the vector mediator mass is $m_V \gtrsim 4 \,\textrm{GeV}$ and the solution to
the $(g-2)_{\mu}$ anomaly does not require any fine tuning of the vector and axial-vector couplings of the $V$ with the muon. Despite a similar mass range for the flavour anomalies in models \textit{44} and \textit{64}, the observables $B_s \to \mu \mu$ and $R_{\Delta M_{s}} $ are sensitive to the momentum dependence of their couplings and thus will be able to discriminate between them. 

FIPs with $m_V > 20\,\textrm{GeV}$ are also likely to provide a good fit to the $b \to s l^+ l^-$ anomalies. However, we have worked here under the assumption that the mixing of $V$ with the electroweak sector of the SM can be neglected. As this  hypothesis may break down when approaching closely the $Z$ mass from below, possibly leading to additional limits from processes such as $Z \to \gamma V$ and others, we leave a proper study of the ``transitional'' regime between light and heavy NP to future work.

While in model~\textit{44} the only viable unified solution lies in the $m_V \gtrsim 4\,\textrm{GeV}$ range, in model~\textit{64} the global analysis of $b\to s l^+ l ^-$ observables highlights additionally a region of good fit at $m_V\approx 0.01-1\,\textrm{GeV}$ passing all flavour constraints. However, this region of the parameter space is characterised by a large muon coupling and requires large fine tuning of the vector and axial-vector contributions to satisfy the $(g-2)_{\mu}$ measurement. 
More importantly, we find that an 
upper bound on the muon coupling imposed by the $\Gamma(W\to \mu\nu V)$ measured width entirely excludes this region of the parameter space. 

In summary, we have highlighted in this work viable solutions for a combined explanation of the $(g-2)_{\mu}$ and $b\to s l^+ l ^-$ anomalies, based on the simple exchange of a light particle and 
characterised by a minimal set of assumptions on the heavy UV completion.
Our goal is that of providing a self-standing and fairly broad 
compendium of the low-energy experimental bounds affecting these scenarios. 
The emerging parameter space regions and coupling strengths can then be used at face value to guide the model-building efforts in the high-energy sector.

\bigskip
\noindent \textbf{Acknowledgments}
\medskip

\noindent
L.D. has been supported in part by the INFN ``Iniziativa Specifica'' Theoretical 
Astroparticle Physics (TAsP-LNF). This project has received funding from the European Union’s Horizon 2020 research and innovation programme under the Marie Sklodowska-Curie grant agreement No. 101028626. The work of M.F. is supported by the Deutsche Forschungsgemeinschaft (DFG, German Research Foundation) under grant  396021762 - TRR 257, ``Particle Physics Phenomenology after the Higgs Discovery''. K.K. is partially supported by the National Science Centre (Poland) under the research Grant No.~2017/26/E/ST2/00470. E.M.S. is supported in part by the National Science Centre (Poland) under the research Grant No.~2017/26/D/ST2/00490. 

\appendix

\section{$B$-meson decays in the DEFT basis\label{app:a}}

This appendix collects the most important ingredients for estimating the FIP contribution to $B$-meson related observables,
as derived directly from the DEFT operators. We eventually compare the results to the standard WET in order to provide a simple dictionary which can be used to re-adapt existing codes to the case of light particles. We use the convention $\sigma_{\mu\nu}=i[\gamma_{\mu},\gamma_{\nu}]/2$ and as customary define $B$ mesons as containing the $b$ quark instead of the antiquark. 

\subsection{$B \to K$ process}\label{app1}

We start with a simple $B \to K \ell \bar \ell$ process. Following the standard procedure, which relies on factorisation of the full amplitude into a hadronic and a leptonic part, one can express the $ \bar B (p) \to \bar K (k) \ell (k_1) \ell (k_2) $ amplitude as~\cite{Bobeth:2007dw}
\begin{align}
    \mathcal{M} ( \bar B \to \bar K \ell \ell) &= \frac{i \, \alpha_{\rm em}}{4 \pi \mathcal{N}} \left( F_V p^\mu [\bar \ell \gamma_\mu \ell] + F_A p^\mu [\bar \ell \gamma_\mu \gamma_5 \ell]\right. \nonumber \\
 & \ \left. +(F_S + \cos \theta F_T) [\bar \ell \ell] + (F_P + \cos \theta F_{T5})  [\bar \ell \gamma_5 \ell ]  \right)\,,
\end{align}
where $\mathcal{N}^{-1}=(4 G_F/\sqrt{2})\,V_{tb}V^{\ast}_{ts}$ and $\theta$ is the angle between the $B$ and the exiting lepton $\ell$  (which is assumed to be a muon in this work) in the $B$-meson frame. We have furthermore denoted with brackets the contraction of the leptonic current on the vacuum di-muon final state, such as for instance,
\begin{align}\label{app:brac_not}
    [\bar \ell \gamma_\mu \gamma_5 \ell] ~\equiv~ \langle \ell^- \ell^+ | \bar \ell \gamma_\mu \gamma_5 \ell | 0 \rangle \,.
\end{align}
The coefficients $F_V$, $F_A$, $F_S$, $F_P$, $F_{T}$ and $F_{T5}$ are given in Ref.~\cite{Bobeth:2007dw} as functions of the Wilson coefficients of the WET basis. 
The $B \to K$ matrix elements of the vector
current are usually parametrised as~\cite{Bailey:2015dka}
\begin{align}
\langle \bar K(k) | \bar s \gamma^{\rho} b | \bar B(p) \rangle &=  f_+(q^2) \left(-\frac{q^{\rho} \left(M_B^2-M_K^2\right)}{q^2}+k^{\rho}+p^{\rho}\right)+\frac{f_0(q^2) \,q^{\rho} \left(M_B^2-M_K^2\right)}{q^2} \, ,
\end{align}
where $q=p-k$ and $f_0(q^2)$,  $f_+(q^2)$
are the form factors. The complete amplitude then follows from contracting the effective vertices corresponding to the DEFT operators of Eqs.~(\ref{eq:bs_oper_4})-(\ref{eq:mu_oper_6}).
One obtains 
\begin{align}
F^\deft_V &= \frac{4 \pi \mathcal{N}}{\alpha_{\textrm{em}} \Pi}\left(  \C^{\mu\mu V}_4 \C^{bsV}_4  + \frac{q^2}{\Lambda^2} \C^\deft \right) f_+(q^2) \,, \label{app:fvdeft}\\
F^\deft_A &=  \frac{4 \pi \mathcal{N}}{\alpha_{\textrm{em}} \Pi}\left(  \Ct^{\mu\mu V}_4 \C^{bsV}_4   + \frac{q^2}{\Lambda^2} \Ct^\deft \right) f_+(q^2) \,,\\ 
F^\deft_P  &= -\frac{4 \pi \mathcal{N}  M_\mu }{\alpha_{\textrm{em} } q^2 \Pi} \left[\left( \Ct^{\mu\mu V}_4 \C^{bsV}_4  - \frac{q^2}{\Lambda^2} \Ct^\deft \right)    f_+(q^2)   \left(M_B^2-M_K^2+q^2\right)  \right. \nonumber \\
 & \left. +  \ \Ct^{\mu\mu V}_4 \C^{bsV}_4 f_0(q^2) \frac{\left(M_B^2-M_K^2\right)  \left(q^2-m_V^2\right)}{m_V^2}   \right], \\
F^\deft_S&= F^\deft_{T} = F^\deft_{T5} = 0 \label{app:ftdeft}\ ,
\end{align}
where the propagator $\Pi$ is defined in Eq.~\eqref{eq:pi} and we have introduced 
\begin{align}
\C^\deft ~\equiv~ \C^{\mu\mu V}_6 \C^{bsV}_4 + \C^{\mu\mu V}_4 \C^{bsV}_6  +\frac{q^2}{\Lambda^2} \C^{\mu\mu V}_6 \C^{bsV}_6  \,,\\
\Ct^\deft ~\equiv~ \Ct^{\mu\mu V}_6 \C^{bsV}_4 + \Ct^{\mu\mu V}_4 \C^{bsV}_6  +\frac{q^2}{\Lambda^2} \Ct^{\mu\mu V}_6 \C^{bsV}_6  \,.
\end{align}
 We can now compare Eqs.~(\ref{app:fvdeft})-(\ref{app:ftdeft}) with the corresponding $F_i$ functions expressed in terms of the WET coefficients, shown, e.g., in Ref.~\cite{Bobeth:2007dw}, to define a dictionary between the DEFT and WET approaches:
\begin{align}
\label{eq:DEFTWET_C9}
C_9^{\mu} &\to  \frac{4 \pi \mathcal{N}}{\alpha_{\textrm{em}} \Pi}\left(  \C^{\mu\mu V}_4 \C^{bsV}_4  + \frac{q^2}{\Lambda^2} \C^\deft \right),  \\
    \label{eq:DEFTWET_C10}
C_{10}^{\mu} &\to  \frac{4 \pi \mathcal{N}}{\alpha_{\textrm{em}} \Pi}\left(  \Ct^{\mu\mu V}_4 \C^{bsV}_4  + \frac{q^2}{\Lambda^2} \Ct^\deft \right),  \\
C_P^{\mu} + C_P^{\mu\prime} &\to -\frac{4 \pi \mathcal{N}}{\alpha_{\textrm{em}} \Pi} \frac{2 M_\mu \left(M_b-M_s\right)}{m_V^2} \left(\Ct^{\mu\mu V}_4 \C^{bsV}_4  + \frac{m_V^2}{\Lambda^2} \Ct^\deft \right) \ .
\end{align}
In the end, we obtain the same results as in Eqs.~(\ref{eq:match_c9})-(\ref{eq:match_cpp}), although the coefficients $C_P^{\mu}$ and $C_P^{\mu\prime}$ arise in a combination that cannot be disentangled in the WET framework for this particular process.

\subsection{$B \to K^*$ process}

A similar comparison for $B$ decaying into a vector meson is relatively more involved, due to the larger phase space of the final-state hadrons. Our strategy will be to estimate the relevant amplitude in the helicity basis, following the conventions of Ref.~\cite{Jager:2012uw}. Employing again the factorisation of the amplitude, one can write
\begin{align}
    \mathcal{M} ( \bar B \to \bar K^* \ell \ell) &= a_{V}^{\mu} [\bar \ell \gamma_\mu \ell] + a_{A}^{\mu} [\bar \ell \gamma_\mu \gamma_5 \ell] + a_S  [\bar \ell \ell] + a_P [\bar \ell \gamma_5 \ell] \nonumber  \\
 & \quad  + a_{TR}^{\mu}  \frac{i}{\sqrt{q^2}} [\bar \ell q^{\nu} \sigma_{\mu \nu} P_R \ell]+ a_{TL}^{\mu}  \frac{i}{\sqrt{q^2}}\bar [\ell q^{\nu} \sigma_{\mu \nu} P_L \ell] \ ,
 \label{eq:BtoKsAmp}
\end{align}
where $a_{V}^{\mu}$, $a_{A}^{\mu}$, $a_S$, $a_P$, $a_{TR}^{\mu}$, $a_{TL}^{\mu}$, are coefficients containing the hadronic part of the amplitude specific 
to the $B\to K^{\ast} ll$ process and  
we have used once more the bracket notation of Eq.~\eqref{app:brac_not} to denote the bracket-contracted leptonic currents. Repeating the same procedure as in Appendix~\ref{app1} and introducing the helicity amplitudes defined in Ref.~\cite{Jager:2012uw}, one finds
\begin{align}
H^\deft_V \equiv \epsilon^*_\rho a_{V}^{\rho} &= 
\frac{\mathcal{N} \tilde{V}_{L \lambda}}{\Pi } \left(  \C^{\mu\mu V}_4 \C^{bsV}_4  + \frac{q^2}{\Lambda^2} \C^\deft \right),\label{app:hvdeft} \\
H^\deft_A \equiv \epsilon^*_\rho a_{A}^{\rho} &= 
\frac{\mathcal{N} \tilde{V}_{L \lambda}}{\Pi } \left(  \Ct^{\mu\mu V}_4 \C^{bsV}_4  + \frac{q^2}{\Lambda^2} \Ct^\deft \right), \\
H^\deft_P \equiv  a_{P} + \frac{2 M_\mu}{q^2} q_\rho a_{A}^{\rho} &= 
- \frac{2 \mathcal{N} M_\mu (M_s \tilde{S}_{L}- M_b \tilde{S}_{R})}{\Pi}
\frac{  \Ct^{\mu\mu V}_4 \C^{bsV}_4  \left(q^2-m_V^2\right) }{ q^2 m_V^2}\label{app:hpdeft}\ ,
\end{align}
where  $\epsilon^*_\rho$ denotes the helicity of $K^*$ and the form factors in the helicity basis, $\tilde{V}_{L \lambda}$, $\tilde{S}_{R}$, $\tilde{S}_{L}$, are defined as in Refs.~\cite{Jager:2012uw,Bharucha:2015bzk,Horgan:2015vla}: 
\begin{align}
    \langle \bar K^* | \bar s \slashed \epsilon^* (\lambda) P_{L} b| \bar B \rangle &=   - i M_B \tilde{V}_{L \lambda}, \\ 
    \langle \bar K^* | \bar s  P_{L(R)} b| \bar B \rangle &=    i M_B \tilde{S}_{L(R)}. 
\end{align}

Finally, one can compare Eqs.~(\ref{app:hvdeft})-(\ref{app:hpdeft}) with the WET results of Ref.~\cite{Jager:2012uw} to obtain the same correspondence between the WET and DEFT coefficients $C_9^{\mu}$ and $C_{10}^{\mu}$ as in Eqs.~\eqref{eq:DEFTWET_C9}-\eqref{eq:DEFTWET_C10}. Interestingly, we are now able to lift the degeneracy between the $C_P^{\mu}$ and $C_P^{\mu\prime}$, obtaining two additional relations,
\begin{align}
\label{eq:DEFTWET2}
C_P^{\mu} &\to -\frac{4 \pi \mathcal{N}}{\alpha_{\textrm{em}} \Pi} \frac{2 M_\mu M_b}{m_V^2} \left(\Ct^{\mu\mu V}_4 \C^{bsV}_4  + \frac{m_V^2}{\Lambda^2} \Ct^\deft \right), \\
C_P^{\mu\prime} &\to \frac{4 \pi \mathcal{N}}{\alpha_{\textrm{em}} \Pi} \frac{2 M_\mu M_s}{m_V^2} \left(\Ct^{\mu\mu V}_4 \C^{bsV}_4  + \frac{m_V^2}{\Lambda^2} \Ct^\deft \right). 
\end{align}
\bigskip

\subsection{$B_s \to \mu \mu$ and $B_s$-mixing processes\label{app:A3}}

For the case of the $B_s \to \mu \mu$ decay,  the kinematics of the process can be derived directly from the matrix elements:
\begin{align}
\langle 0 | \bar s \gamma^\rho \gamma^5 b  | \bar B_s(p) \rangle &=  i f_{B_s}p^\rho\ , \\
\langle 0 | \bar s \gamma^5 b | \bar B_s(p) \rangle &=  - i \frac{M_{B_s}^2}{M_b+M_s}f_{B_s}  \ ,
\end{align}
where $f_{B_s}$ is the decay constant of the $B_s$ meson.

As was pointed out in Sec.~\ref{sec:BtoK}, only the DEFT operators of dimension four contribute to this process, leading to
\begin{align}
\mathcal{M} (\bar B_s \to \mu \mu) &= - \frac{M_\mu f_{B_s}}{\Pi\,  m_V^2} \Ct^{\mu\mu V}_4 \C^{bsV}_4  \left(M_{B_s}^2-m_V^2\right) \,  [\bar \ell \gamma_5 \ell] \ .
\end{align}
The above can be compared with the SM prediction to obtain the ratio defined in Eq.~\eqref{eq:bsmu_exp}.  One can then calculate the $R_{B_s} $ ratio as
\begin{align}
R_{B_s}  &= \frac{ \textrm{BR}(B_s\to \mu^+\mu^-)_{\textrm{exp.}} }{  \textrm{BR}(B_s\to \mu^+\mu^-)_{\textrm{SM}}} - 1 \nonumber \\
&= \left| 1+ \frac{4 \pi\,  \C_4^{bsV} \Ct_4^{\mu\mu V}  \left(m_V^2-\mBs^2 \right)}{C_{10}^{\rm SM}  \Pi\,  \alpha_{\textrm{em}} m_V^2 } \right|^2 -1 \ ,
\end{align}
and check using the WET expression~\cite{Beneke:2017vpq} that the link between the DEFT and WET operators can be obtained from the $B \to K$ one, as advertised in the main text. Note that the amplitude is suppressed when the FIP mass $m_V$ nears $M_{B_s}$. By expanding in the small coupling limit, one then recovers the main text result Eq.~\eqref{eq:RBs44}.

For the case of $B_s$ mixing, the relevant four-quark currents correspond directly to the operators of the WET basis with colour indices contracted in pairs (we follow the conventions of Ref.~\cite{FermilabLattice:2016ipl}): 
\begin{align}
    \mathcal{O}_1 &= (\bar s \gamma^\mu P_L b) \, (\bar s \gamma^\mu P_L b)\ , \\
    \widetilde{\mathcal{O}}_1 &= (\bar s \gamma^\mu P_R b) \, (\bar s \gamma^\mu P_R b)\ ,\\
    \mathcal{O}_2 &= (\bar s P_L b) \, (\bar s P_L b)\ ,\\
    \widetilde{\mathcal{O}}_2 &= (\bar s P_R b) \, (\bar s P_R b)\ , \\
    \mathcal{O}_4 &= (\bar s P_L b) \, (\bar s P_R b) \ .
\end{align}
After simplifying the effective vertices, one obtains 
\begin{align}
    C_1 &\to  -\frac{1}{\Pi } \left((\C^{bsV}_4)^2 + \frac{\C^\deft_{B_s} M_{B_s}^2}{\Lambda ^2} \right),\label{eq:bsmix1} \\
    \widetilde{C}_1 &\to 0 \, ,\label{eq:bsmix1t}\\
      C_2 &\to  - \frac{M_b^2}{ m_V^2 \Pi } \left( (\C^{bsV}_4)^2 + \frac{\C^\deft_{B_s} M_{B_s}^2}{\Lambda ^2} \right),\label{eq:bsmix2}\\
     \widetilde{C}_2 &\to  -\frac{M_s^2}{ m_V^2 \Pi }  \left( (\C^{bsV}_4)^2+ \frac{\C^\deft_{B_s} M_{B_s}^2}{\Lambda ^2} \right),\label{eq:bsmix3}
    \end{align}
\begin{align}
     C_4  &\to \frac{2 M_b M_s }{ m_V^2 \Pi } \left( (\C^{bsV}_4)^2 + \frac{\C^\deft_{B_s} M_{B_s}^2}{\Lambda ^2} \right),\label{eq:bsmix4}
\end{align}
where we have defined
\begin{equation}
\C^\deft_{B_s} = \C^{bsV}_4 \C^{bsV}_6 +  \frac{M_{B_s}^2}{\Lambda^2} (\C^{bsV}_6)^2 .
\end{equation}

Eventually, the $B_s$-mixing matrix element is computed  
in terms of bag parameters $B^{(i)}_{B_s}$, relative to each operator $\mathcal{O}_i$, by estimating the matrix element of each four-quark operator. We define~\cite{FermilabLattice:2016ipl}, 
\begin{align}
    \langle  B_s |\mathcal{O}_1  |  \bar B_s \rangle &=   \langle  B_s |\widetilde{\mathcal{O}}_1  | \bar B_s \rangle = \frac{2}{3} f_{B_s}^2 \mBs^2  B^{(1)}_{B_s}\ , \\ 
    \langle B_s |\mathcal{O}_2  | \bar B_s \rangle &=   \langle  B_s |\widetilde{\mathcal{O}}_2  | \bar B_s \rangle = -\frac{5}{12} \left( \frac{\mBs}{M_b+M_s}\right)^2 f_{B_s}^2 \mBs^2  B^{(2)}_{B_s}\ , \\ 
      \langle B_s |\mathcal{O}_4  | \bar B_s \rangle &= \frac{1}{2} \left[\left( \frac{\mBs}{M_b+M_s}\right)^2+\frac{1}{6} \right]f_{B_s}^2 \mBs^2  B^{(4)}_{B_s} \ .
\end{align}
We would like to stress that, compared to the standard WET approach, the bag parameters should not be run from a high scale since the amplitude of the process is estimated directly at the $B_s$ scale. Note also that the DEFT coefficients used in our results do not run, as the currents charged under colour are (axial)-vector. 

The experimental observable of interest is the ratio $R_{\Delta M_s}$ between the SM-predicted and NP contribution to the mass difference of the neutral $B_s$ mesons~\cite{Zyla:2020zbs}:
\begin{equation}
 R_{\Delta M_s}= \left| 1+ \sum_{i=1}^2 R_i \frac{C_i+\widetilde{C}_i }{C_{1}^{\rm SM}(\mu_b) } + R_4 \frac{C_4}{C_{1}^{\rm SM}(\mu_b) }  \right|-1 \ ,
\end{equation}
where we have defined the ratios of bag parameters at the $b$ scale by
\begin{align}
R_i \, \equiv \, \frac{\langle B_s| \mathcal{O}_i(\mu_b)| \bar{B}_s \rangle}{\langle B_s| \mathcal{O}_1(\mu_b)| \bar{B}_s \rangle }\ .
\end{align}
In the  small coupling approximation, one finally finds
\begin{equation}
  R_{\Delta M_s}  \simeq - \frac{\, (\mathcal{C}_4^{bsV})^2  }{C_{1}^{\rm SM}(\mu_b) \, m_V^2}  \frac{(\mBs^2-m_V^2)(m_V^2+R_2 M_b^2)}{(m_V^2-\mBs^2)^2 + \mBs^2 \Gamma_V^2} \,,
\end{equation}
and
\begin{equation}
  R_{\Delta M_s}  \simeq -  \frac{\, \mBs^2 (\mathcal{C}_6^{bsV})^2  }{\Lambda^4  \, C_{1}^{\rm SM}(\mu_b) }  \frac{(\mBs^2-m_V^2)(\mBs^2+R_2 M_b^2)}{(m_V^2-\mBs^2)^2 + \mBs^2 \Gamma_V^2} \,.
\end{equation}

\vspace{0.5cm}               
\bibliographystyle{JHEP}
\bibliography{bibliography}

\end{document}